\documentclass[]{aa}
\usepackage[varg]{txfonts}
\usepackage{epstopdf}
\usepackage{supertabular,booktabs}
\usepackage{multicol}
\usepackage{longtable}
\usepackage{natbib}
\usepackage{rotating}
\usepackage{psbox}
\usepackage{graphicx}
\usepackage{rotating}
\usepackage{pdfpages}
\usepackage{multirow}
\usepackage{graphicx}
\usepackage{caption}
\usepackage{verbatim}
\usepackage{url}
\usepackage{tikz}
\usetikzlibrary{positioning}
\usepackage{amssymb}
\usepackage{booktabs}

\bibpunct{(}{)}{;}{a}{}{,} 

\usepackage[flushleft]{threeparttable}

\usepackage{caption}
\usepackage{xcolor}

\newcommand{\lya}{\mbox{${\rm Ly}\alpha$}}

\newcommand{\zabs}{\ensuremath{z_{\rm abs}}}

\newcommand{\CI}{\ion{C}{i}}

\newcommand{\CIV}{\ion{C}{iv}}

\newcommand{\FeII}{\ion{Fe}{ii}}
\newcommand{\HI}{\ion{H}{i}}

\newcommand{\OI}{\ion{O}{i}}

\newcommand{\PII}{\ion{P}{ii}}

\newcommand{\SII}{\ion{S}{ii}}
\newcommand{\SiII}{\ion{Si}{ii}}

\newcommand{\avg}[1]{\left< #1 \right>} 
\newcommand{\kms}{\ensuremath{{\rm km\,s^{-1}}}}
\newcommand{\cmsq}{\ensuremath{{\rm cm}^{-2}}}

\begin{document}

\title{Molecular gas and star formation in an absorption-selected galaxy: Hitting the bull's eye at $z\simeq 2.46$\thanks{Based on observations performed with the Very Large Telescope of the European Southern Observatory under Prog.~ID 095.A-0224(A).}}
\titlerunning{\HI, H$_2$ and star formation in the main disc of a DLA galaxy at $z\simeq 2.46$}

\author{A.~Ranjan\inst{1}
  \and P.~Noterdaeme\inst{1}
  \and J.-K.~Krogager\inst{1}
  \and P.~Petitjean\inst{1}
  \and S.~A.~Balashev\inst{2}
  \and S.~Bialy\inst{3}
  \and R.~Srianand\inst{4}
  \and N.~Gupta\inst{4} 
  \and J.~P.~U.~Fynbo\inst{5}
  \and C.~Ledoux\inst{6}
\and P.~Laursen\inst{7}
  }

\institute{ 
Institut d'astrophysique de Paris, UMR\,7095, CNRS-SU, 98bis bd Arago, 75014 Paris, France
\and Ioffe Institute of RAS, Polytekhnicheskaya 26, 194021 Saint-Petersburg, Russia 
  \and Raymond and Beverly Sackler School of Physics \& Astronomy, Tel Aviv University, Ramat Aviv 69978, Israel
  \and Inter-University Centre for Astronomy and Astrophysics, Post Bag 4, Ganeshkhind, 411 007, Pune, India
\and Cosmic Dawn Center, Niels Bohr Institute, Copenhagen University, Juliane Maries Vej 30, 2100 Copenhagen Ø, Denmark
\and European Southern Observatory, Alonso de C\'ordova 3107, Casilla 19001, Vitacura, Santiago, Chile
\and Oskar Klein Centre, Dept. of Astronomy, Stockholm University, SE-10691 AlbaNova, Stockholm, Sweden
}

\date{Received 17 August 2017}

\abstract{We present the detection and detailed analysis of a diffuse molecular cloud at $z_{\rm abs}=2.4636$ towards the quasar SDSS J\,1513+0352 ($z_{\rm em}\simeq\,2.68$) observed with the X-shooter spectrograph at the Very Large Telescope. We measure very high column densities of atomic and molecular hydrogen, with $\log N(\HI,{\rm H_2}) \simeq 21.8, 21.3$. 
This is the highest H$_2$ column density ever measured in an intervening damped Lyman-$\alpha$ system but we do not detect CO, implying $\log N($CO$)/N$(H$_2)<-7.8$, which could be due to a low metallicity of the cloud.  
From the metal absorption lines, we derive the metallicity to be $Z\simeq 0.15 Z_{\rm \odot}$ and determine the amount of dust by measuring the induced extinction of the background quasar light, $A_{\rm V}\simeq 0.4$. 
We simultaneously detect \lya\ emission at the same redshift, with a centroid located at a most probable impact parameter of only $\rho \simeq 1.4$~kpc. 

We argue that the line of sight is therefore likely passing through the interstellar medium of a galaxy, as opposed to the circumgalactic medium.
The relation between the surface density of gas and that of star formation seems to follow the global empirical relation derived in the nearby Universe although our constraints on the star formation rate and on the galaxy extent remain too loose to be conclusive. We study the transition from atomic to molecular hydrogen using a theoretical description based on the microphysics of molecular hydrogen. We use the derived chemical properties of the cloud and physical conditions ($T_k\simeq 90$~K and $n\simeq 250$~cm$^{-3}$) derived through the excitation of H$_2$ rotational levels and neutral carbon fine structure transitions to constrain the fundamental parameters that govern this transition. By comparing the theoretical and observed \HI column densities, we are able to bring an independent constraint on the incident UV flux, which we find to be in agreement with that estimated from the observed star formation rate.}

\maketitle

\section{Introduction}
Stars are known to form from the gravitational collapse of cold dense gas clumps 
in the interstellar medium \citep[see][]{Leroy2008}. Observations in our own Galaxy and in the nearby Universe have 
further showed that this dense gas is in molecular form (H$_2$). The fact that the gas giving birth to stars is molecular is well known. Although the exact reason for the same is not trivial since H$_2$ is not a good coolant of the interstellar medium (ISM). However, it is energetically favoured compared to atomic hydrogen. In fact, the molecular state of the gas is more the consequence 
of physical conditions favouring both the conversion of atomic into molecular gas and the formation of stars, than a true pre-requisite for star formation (see discussions in 
\citealt{Glover2012}). 
Indeed, it may also be possible that, below a certain metallicity, the gas collapses into stars on a time-scale shorter than needed to form H$_2$ \citep[e.g.][]{Krumholz2012,Michalowski2015}.

Therefore, while investigating globally the relations between atomic gas, molecular gas and star formation at different epochs should bring important clues to understand the formation and evolution of galaxies, it is equally important to understand the micro-physics at play in the gas (i.e. at small scale) for a complete theory of star formation. 

An empirical relation between the surface density of neutral gas and that of star formation is well established over galactic-scales in the nearby Universe \citep[see][]{KennicuttandEvans2012}. It seems 
that this relation actually implicitly results from a strong relation between H$_2$ and star formation seen on kpc or sub-kpc scales \citep[e.g.][]{Schruba2011} and a conversion from \HI\ to H$_2$. This conversion can be described phenomenologically 
as a relation between molecular fraction and hydrostatic pressure at galactic mid-plane \citep{Blitz2006}, but 
also from the first principles as an equilibrium between formation and destruction of the molecule. This has 
been intensively studied theoretically and observationally in the local interstellar medium (see review by \citealt{Wakelam2017}). In short, the formation of H$_2$ in cold neutral medium is dominated by three-body reactions on the surface of dust grains \citep[e.g.][]{Hollenbach1971}, even at relatively low-metallicities. The gas-phase production of H$_2$ is not efficient although it is the only route to form H$_2$ in pristine gas \citep[e.g.][]{Black1987}. 
H$_2$ is destroyed primarily through absorption of UV photons in the Lyman-Werner (LW) bands of H$_2$, followed by a 
radiative decay onto a dissociative state in approximately 12\% of the cases \citep{Abgrall1992}. Therefore, shielding of 
H$_2$ occurs both as self-shielding (in these LW bands) and shielding from dust which 
absorbs the incident UV radiation over an extended wavelength range.

Large and deep surveys now permit the study of star formation properties of galaxies in the distant Universe \citep[see e.g.][]{Madau2014}. Furthermore, thanks to the advances of sub-mm astronomy, it is also possible to detect large molecular reservoirs traced by CO emission lines \citep[see][and references therein]{Carilli2013}. Finally, the forthcoming Square Kilometer Array \citep{SKA} will open the possibility to study the atomic gas in emission beyond the local Universe. 

Currently the only way to study the small-scale chemical and physical properties of neutral gas is by observing the absorption lines imprinted by the gas on the spectra of background sources, such as quasars. Absorption spectroscopy indeed allows one to study several gas phases simultaneously while probing the associated galaxy irrespective of its brightness. Hence, this technique efficiently traces the overall population of galaxies.

However, gas phase studies remain quite disconnected from that of the star formation and macroscopic properties of galaxies. 
This is due not only to the difficulty in detecting in emission the galaxy counterpart of the absorbing gas (or finding background sources behind emission-selected galaxies),
but also because the cross-section selection makes most lines of sight only probe the very outskirts of these galaxies. For the highest column density absorbers (the so-called damped Lyman-$\alpha$ absorbers [DLAs], see \citealt{wolfe2005}), the typical impact parameters are of the order $\rho\sim10$~kpc or less \citep{RahmatiandSchaye2014, Rubin2015, Krogager+17}. 
Improved observational strategies and new powerful instruments have now permitted the detection of the host galaxies responsible for an increasing number of DLAs \citep[e.g.][]{Fynbo2010,Fynbo2011,Noterdaeme2012, Krogager2013, Hartoog2015, Srianand2016, Ma2018, Neeleman2018}. Statistical relations start to emerge, such as a relation between the metallicity of the gas probed in absorption and the impact parameter and luminosity of the host galaxy \citep{Krogager+17}. 
However, the number of detections remains low and, it is crucial to study lines of sight that pass through the {\sl interstellar} medium of high-redshift galaxies which, as its name indicates, refers to the gas 
collocated with stars in galaxies.
Since the gas falls in the gravitational potential of galaxies, it is naturally expected that higher column densities of gas will trace more central regions, where stars are found. This results in an anti-correlation between \HI\ column density and impact parameter, predicted by simulations \citep[e.g.][]{Pontzen2008} and observed over a wide range of redshifts  \citep{zwaan2005,Rao2011,Krogager+12}.
Based on these considerations, \citet{noterdaeme2014} searched for extremely strong DLAs (ESDLAs, with $\log N(\HI)>21.7$) in the Sloan Digital Sky Survey \citep{York2000}. 

They statistically showed that these systems are connected to the population of Lyman-$\alpha$ emitters and
that the impact parameters are likely small. Simulations furthermore show that the high end of the $N(\HI)$ distribution is sensitive to the stellar feedback and the formation of molecules \citep[e.g.][]{Altay+13, Bird2014}. We have, therefore, initiated an observational campaign to search not only for the associated galaxies in emission but also for the molecular gas in absorption. 
We note two interesting cases here, the ESDLAs at $\zabs=2.207$ and $\zabs=2.786$ towards SDSS\,J1135$-$0010 and SDSS\,J0843$+$0221, respectively. In the former case, a star-forming galaxy is detected with its centroid located 
at a very small impact parameter $\rho<1$~kpc, directly showing that the line of sight passes through 
the main star-forming region of the galaxy \citep{Noterdaeme2012}. Unfortunately, due to a higher redshift Lyman limit system, H$_2$ cannot be detected. In the second case, extremely strong H$_2$ lines are detected \citep{Balashev+17} but the galaxy emission remains undetected. Due to the low metallicity in the latter case, we can expect the host galaxy to be of low mass and low 
luminosity \citep[e.g.][]{Fynbo2008, Krogager+17}.
Here we present the simultaneous detection of molecular gas with very high H$_2$ column density and star-formation activity from a galaxy at $z \approx 2.46$ aligned with 
the background quasar SDSS J\,151349.52$+$035211.68 (hereafter J\,1513$+$0352).

Standard $\Lambda$CDM flat cosmology is used for the paper with $\rm H_0$ = 67.8 $\pm$ 0.9 km$\rm s^{-1}Mpc^{-1}$, $\rm \Omega_{\Lambda}$ = 0.692 $\pm$ 0.012 and $\rm \Omega_{m}$ = 0.308 $\pm$ 0.012 \citep{Planck2016}.
The details of the observations and data reduction are presented in Section~\ref{observations}. The analysis of the chemical properties of the gas and the dust content are described in Sect.~\ref{analysis}. The detection and properties of the associated galaxy is presented 
in Sect.~\ref{galaxy}. We discuss and compare our results with global relationships between gas and star formation and discuss the gas microphysics in Sect.~\ref{Discussion} before finally summarizing our findings and concluding in Sect.~\ref{Conclusion}.

\section{Observations and data reduction \label{observations} }
The quasar J\,1513$+$0352 was observed on four different nights in service mode, on 15-16 Apr 2015 and 14-15 May 2015 under ESO program ID 095.A-0224(A) with the multi-wavelength, medium resolution spectrograph X-shooter mounted at the Cassegrain focus of the Very Large Telescope UT2. The observations were performed under good conditions (average seeing $\sim\,0.69$ and airmass $\sim 1.15$). 
Each time, the target was observed in a nodding sequence totalling 2960\,s ($2\times1480$\,s) in the UVB, 2860\,s ($2\times1430$\,s) in the VIS and 2880\,s ($2\times3\times480$\,s) in the NIR. The slit widths used were 1.6, 0.9 and 1.2 arcsec for the UVB, VIS and NIR arm respectively. All observations were performed with the slit aligned with the parallactic angle, which changed little between the observations ($\pm 10^{\circ}$). The log of the observations is summarised in Table~\ref{journal_of_observations}. 
We reduced the data using the X-shooter pipeline \citep{Modigliani2010} and combined individual exposures weighting each pixel by the inverse of its variance to obtain the final 2D and 1D spectra used in this paper.  The typical signal-to-noise (per pixel) ratios achieved are $S/N \simeq 25, 23$ and  15 at 4700, 6500 and 12400~{\AA}, respectively.
Since the seeing was significantly smaller than the slit width, in particular in the UVB, the spectral resolution achieved is higher than the nominal resolution for the given slit widths. This is in agreement with the resolution expected in cases where the seeing is significantly smaller than the slit width as in \citet{Fynbo2011}. Moreover, we checked the actual spectral resolution directly from the combined spectra by first fitting the numerous \FeII\ lines in the VIS part of the spectra that was obtained with a narrower slit using a multi-component model, with the column density, Doppler parameter, redshift and resolution as free parameters. We found a spectral resolution of $R$=9300, which is consistently slightly higher than the nominal resolution 
for the given slit width. We then used this model to derive the resolution in the UVB region by using the \ion{Fe}{ii}~$\lambda1608$ and \ion{Fe}{ii}~$\lambda1611$ lines
and obtained $R$=7300. This was confirmed by fitting the \ion{Si}{ii} lines along with the \ion{Fe}{ii} lines.

\begin{table*}
    \caption{Journal of observations}
    \label{journal_of_observations}
    \centering
     \begin{tabular}{ c c c c c c }
        \hline \hline
        observing date & slit widths & exposure time  & Seeing & Airmass & Position Angle\\    
                       & (arcsec)    & (s)            &  (arcsec)   &       &    (degrees)  \\
        \hline
        15 Apr 2015    & 1.6(UVB) & 2$\times$1480 & 0.9 & 1.14 & -179 \\
        15 Apr 2015    & 0.9(VIS) & 2$\times$1430 & 0.9 & 1.14 & -179 \\
        15 Apr 2015    & 1.2(NIR) & 2$\times$(3$\times$480)  & 0.9 & 1.14 & -179 \\
               
        16 Apr 2015    & 1.6(UVB) & 2$\times$1480 & 0.7 & 1.16 & -160 \\
        16 Apr 2015    & 0.9(VIS) & 2$\times$1430 & 0.7 & 1.16 & -160 \\
        16 Apr 2015    & 1.2(NIR) & 2$\times$(3$\times$480)  & 0.7 & 1.16 & -160 \\        
        
        14 May 2015    & 1.6(UVB) & 2$\times$1480 & 0.6 & 1.14 & -172 \\
        14 May 2015    & 0.9(VIS) & 2$\times$1430 & 0.6 & 1.14 & -172 \\
        14 May 2015    & 1.2(NIR) & 2$\times$(3$\times$480) & 0.6 & 1.14 & -172 \\        
        
        15 May 2015    & 1.6(UVB) & 2$\times$1480 & 0.6 & 1.15 & -164 \\
        15 May 2015    & 0.9(VIS) & 2$\times$1430 & 0.6 & 1.15 & -164 \\
        15 May 2015    & 1.2(NIR) & 2$\times$(3$\times$480)  & 0.6 & 1.15 & -164 \\

        \hline 
     \end{tabular}
      \tablefoot{
      The three different values for slit widths and exposure times correspond to the UVB, VIS and NIR arms, respectively. The exposure time for NIR data is sub-divided into 3 integrations (NDIT). Here, 'Seeing' refers to the "delivered seeing corrected by airmass" in the VIS Band and the value of 'Airmass' is recorded at start of exposure.
      The last column gives the position angle at the start of the observations.}
\end{table*}

\section{Absorption analysis\label{analysis} }
\begin{figure}
\centering
   \includegraphics[width = 1.0\hsize]{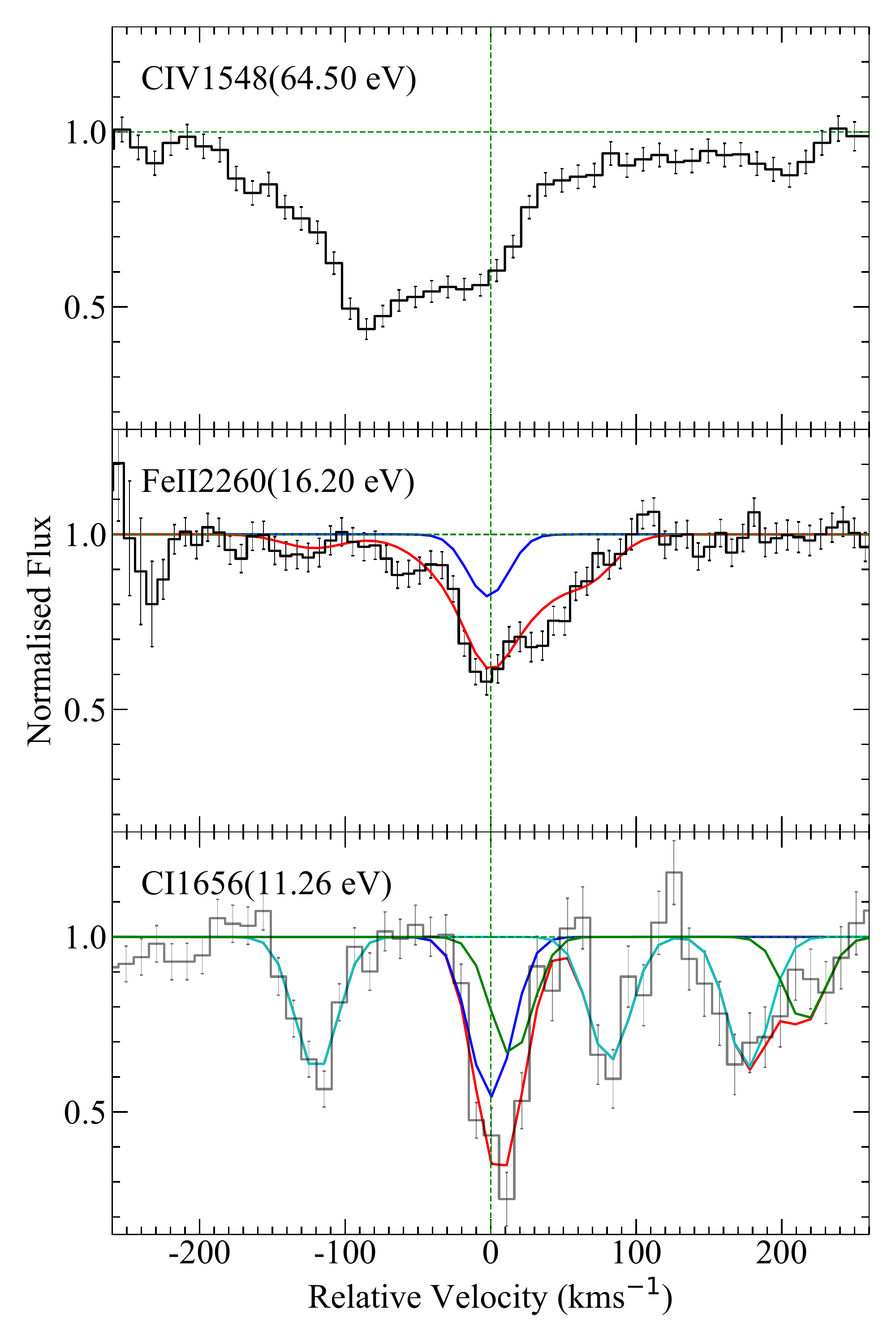}
   
   \caption{Velocity profiles of selected species associated to the DLA system at  $\rm z=2.4636$ towards J\,1513$+$0352, tracing the ionised gas (\CIV), neutral atomic gas (\FeII) and molecular gas (\CI). Note that despite the apparent complexity, the \CI\ profile is actually characterised by a single component (shown in blue) and other absorption seen in the panel arise from excited fine-structure levels (J=1 and J=2, in cyan and green, respectively, with the total profile in red). The blue line in the \FeII\ panel shows the contribution from the $v=0$ component.}
     \label{Metal-vel}
\end{figure}

The DLA at $z_{\rm abs}=2.4636$ towards the quasar SDSS J\,1513$+$0352 presents extremely strong absorption lines from atomic (\HI) and molecular (H$\rm _2$) hydrogen as well as absorption lines from metals in different ionization stages. The kinematic profiles of selected species tracing different phases are illustrated in Fig.~\ref{Metal-vel}. 
As in the overall population of DLAs, singly-ionised metal species are seen and are expected to be in their dominant ionisation stage in the neutral gas, where only photons with energies less than 13.6~eV can penetrate. The corresponding profile is compact with a main central component that likely contains the bulk of the atomic gas. A single component of neutral carbon is seen at the velocity of this main metal component. Since the first ionisation potential of carbon is lower than the energy of photons that destroy H$_2$ \citep[see e.g.][]{Noterdaeme17}, this component also likely contains most of the molecular hydrogen. In turn, high ionisation species trace ionised gas and have a very different kinematic profile. Indeed, the \CIV\ profile is strongest where there is no component seen in the neutral or low ionisation species, possibly indicative of outflowing material \citep{Fox2007}. Since this ionised phase is, apparently, physically and kinematically disconnected from the neutral and molecular phases, it is not considered further in this paper, as we concentrate on the properties of the neutral and molecular gas. 

We analysed the absorption line profiles of \HI, H$\rm _2$ and low-ionization metals using standard multiple Voigt profile fitting using VPFIT v10.2  \citep{Carswell2014}. The gas-phase abundances are expressed relative 
to the Solar values as:

\begin{equation}
\rm {[X/H]} = \log\left(\frac{N({X})}{N({H_{tot}})}\right) - \log\left(\frac{X}{H}\right)_{\odot},
\end{equation}

\noindent where $\rm N({H_{tot}})=N(\HI)+2N(H_2)$ is the total hydrogen column density, $\rm X$ denotes the metal species and $\rm (X/H)_{\odot}$ is the solar metal abundance. We adopted the same Solar abundances as in \citet{DeCia2016}, i.e. from \citet{Asplund+09}, following the recommendations of \citet{lodders2009} about the choice of photospheric, meteoritic or average values. 

\subsection{Neutral carbon \label{neutral_carbon} }

\begin{figure}
\centering
   \includegraphics[width = 0.5\textwidth]{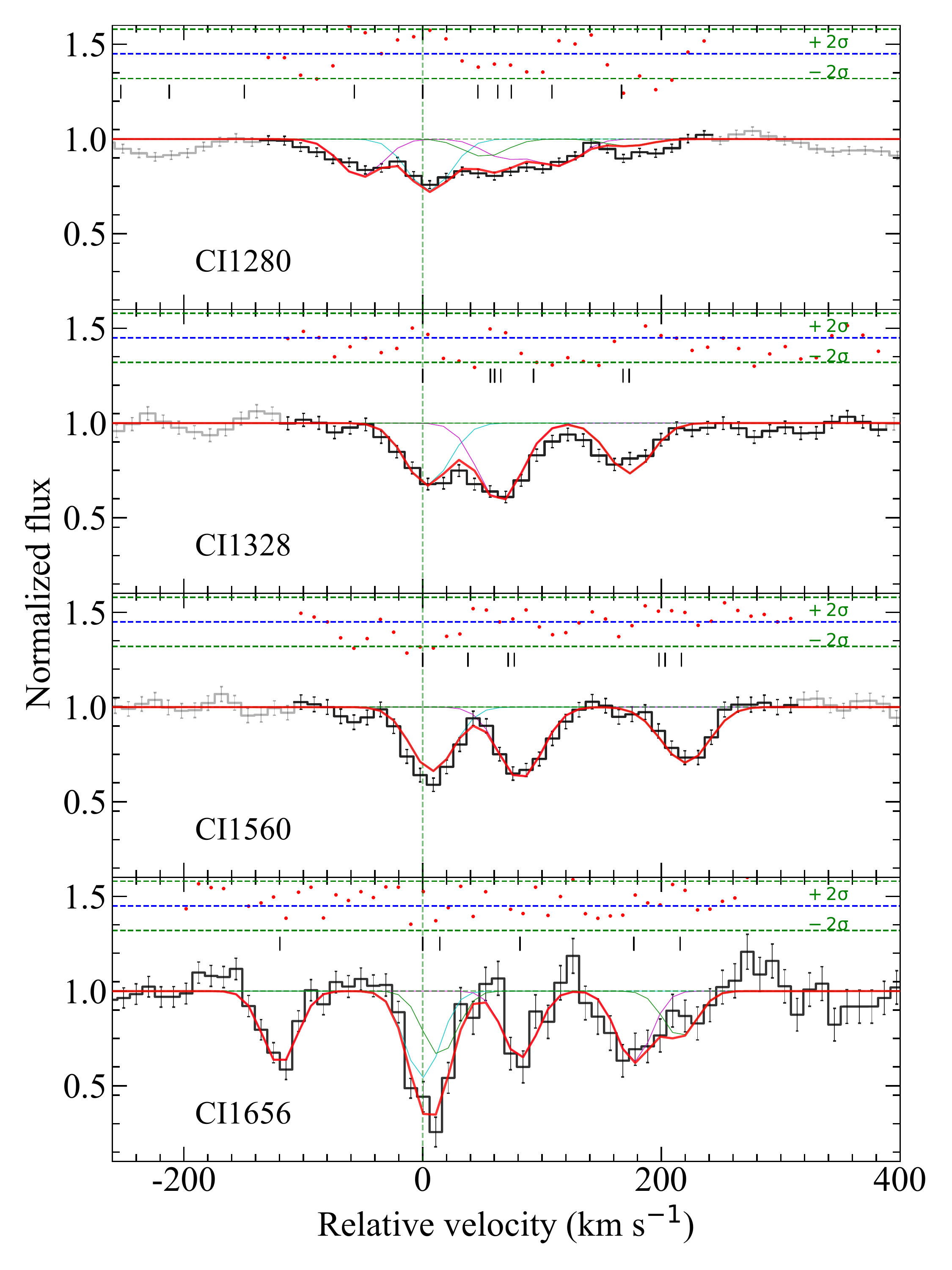}
     \caption{Fit to the \ion{C}{I} lines associated with the DLA. The panels show the 4 lines used for fitting. The data is represented in black and the best-fit synthetic profile in red with residuals in a $\pm 2\,\sigma$-scale on the top of each panel. The individual contributions from $J=0$ (\ion{C}{i}), $J=1$ (\ion{C}{i}*) and $J=2$ (\ion{C}{i}**) are displayed with cyan, magenta and green lines respectively.      \label{CI_profile}}
\end{figure}

Neutral carbon is detected 
in the fine-structure levels of the electronic ground state denoted 
here \ion{C}{i} (J=0), \ion{C}{i*} (J=1) and \ion{C}{i**} (J=2). We derived the column density in each fine-structure level by 
simultaneously fitting the \CI\ bands located at $\lambda = 1280, 1328, 1560$ and 1656~{\AA} in the DLA rest-frame (see Fig.~\ref{CI_profile}). The single component is fitted assuming the same redshift and 
Doppler parameter for all fine-structure levels. We obtain a Doppler parameter $b=3.9 \pm 0.3$~\kms and column densities of $\log N(\CI,J=0,1,2)= 14.82\pm0.18$, $14.60\pm0.06$ and $14.03\pm0.06$, respectively\footnote{Here, and in the following, column densities expressed in log are in units of cm$^{-2}$, i.e., $\log N$ stands for $\log N/$cm$^{2}$.}. 
The best-fit synthetic profile is overplotted on the data in Fig.~\ref{CI_profile}. 
We note here that the bluer \CI\ bands are detected in the UVB but not used to constrain the fit since they are partly blended with the Lyman-$\alpha$ forest. We, however, checked that the calculated profile using our derived best-fit parameter values is consistent with the data in these regions as well.  
For the remaining figures in the paper, where we show the absorption profile as a function of relative velocity, the zero point of the velocity scale ($v=0$~km\,s$^{-1}$) corresponds to the \ion{C}{i} component at $z = 2.46362$. 

\subsection{Neutral and molecular hydrogen }
\label{neutral_and_molecular_hydrogen}

\begin{figure*}
\centering
   \includegraphics[trim={0.5cm 6.5cm 0.5cm 2.5cm},clip,width=\hsize]{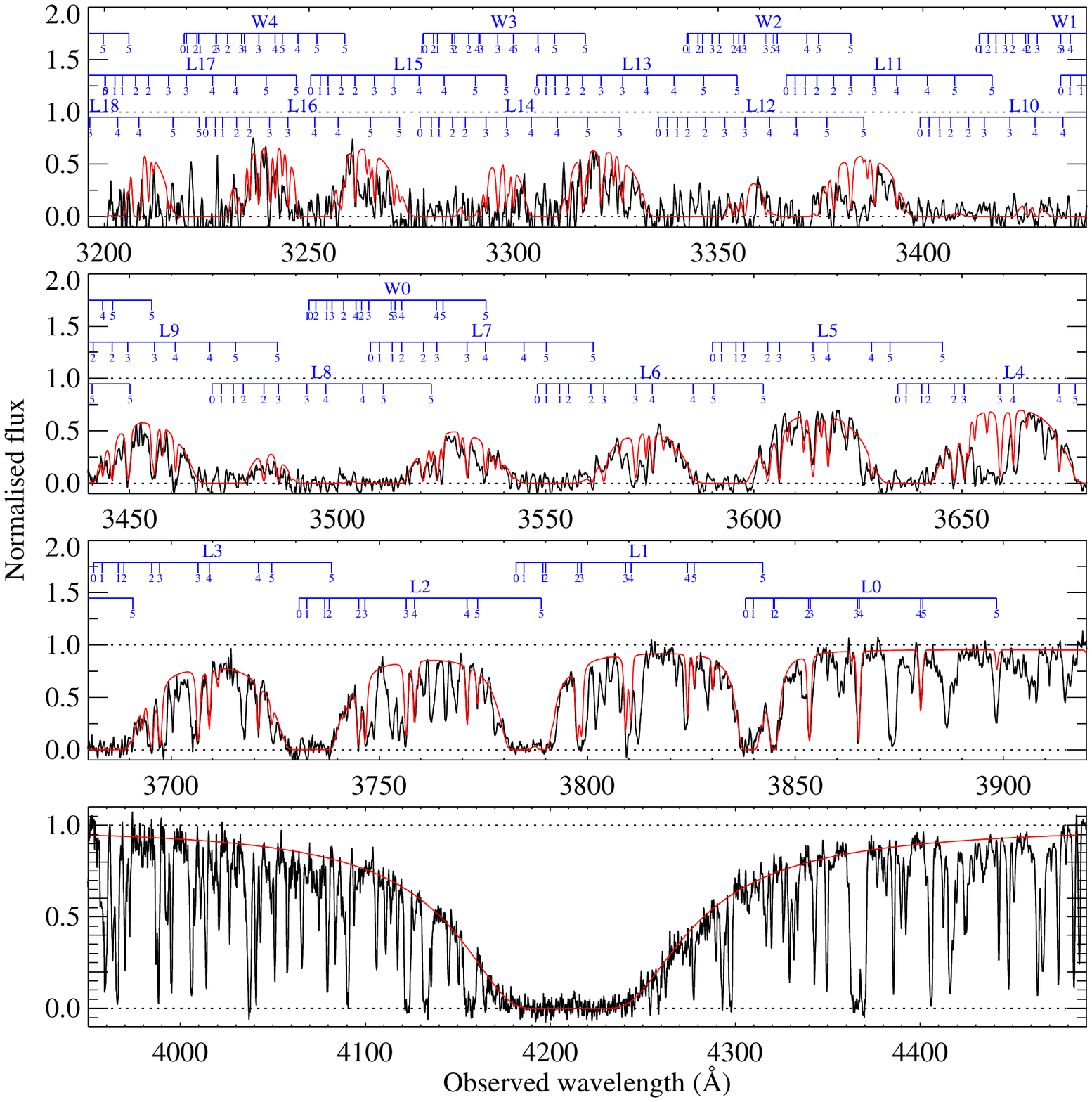}
\caption{Portion of the X-shooter spectrum of J\,1513$+$0352 covering the absorption lines of \HI\ (Lyman-$\rm \alpha$ shown in the bottom panel) and H$\rm _2$ (top three panels) from the DLA at $z$=2.4636. The normalized data are shown in black and the synthetic profile is over-plotted in red. We note the very strong flux suppression at the bluest wavelengths due to the increased overlapping of H$\rm _2$ lines. Horizontal blue segments connect rotational levels (short tick marks) from a given Lyman (L) or Werner (W) band, as labelled above. \label{HI_and_H2}}
\end{figure*}

Figure~\ref{HI_and_H2} shows the portion of the UVB spectrum with absorption originating from the Lyman series of \HI\ as well as the Lyman and Werner bands of H$\rm _2$ associated with the DLA. Because the molecular hydrogen lines are numerous and damped, they are blended with each other, significantly decreasing the apparent continuum. In order to retrieve the true unabsorbed quasar continuum, we used the same approach as in \citet{Balashev+17}. We first, simultaneously, fitted the damped \HI\ and H$\rm _2$ lines together with the continuum using the full wavelength range from 3200 to 4500~{\AA}. We found that the first 10 Chebyshev polynomials were sufficient to model the unabsorbed continuum. We then used this derived continuum to perform the complete fit of \HI\ and H$\rm _2$ lines using VPFIT and removing the regions apparently contaminated by  unrelated absorption (e.g. \lya\ forest).

We obtain a total \HI\ column density of $\rm \log N(\HI)=21.83 \pm 0.01$, which confirms the value derived  by \citet{noterdaeme2014} from the low S/N and low resolution SDSS spectrum ($\rm \log N(\HI)\simeq 21.80$).
The total H$\rm _2$ column density is found to be $\rm \log N(\rm H_2)= 21.31 \pm 0.01$. This is the highest value reported to date in an intervening DLA, much above typical H$_2$ columns (	$\log N(\rm H_2)\,\sim\,18.2$) seen in other DLAs \citep{noterdaeme2008} although only slightly higher than the recent detection associated to another extremely strong DLA recently reported by \citet{Balashev+17}.
The total hydrogen column density is then also amongst the highest seen in DLAs, $\rm \log N(H_{tot}) = 22.04 \pm 0.01$, with a very significant {\sl overall} molecular fraction of $f$ = $\rm 2N({H_2})/(2N({H_2})+N(\HI)) = 0.38\pm0.02$. Since low-ionisation metals suggest that atomic gas is present in different components, this represents a lower limit to the molecular fraction in the H$_2$ bearing cloud (likely present in a single component, as suggested from the \CI\ profile). Furthermore, at such high hydrogen column density, the gas is likely to be fully molecular in the cloud core with the atomic gas corresponding to its external layers. We will discuss this in details later in Sect.~\ref{Bialys_formalism}. 

As for other known H$\rm _2$ systems, most of the total H$\rm _2$ is found in the first two rotational levels (see Table~\ref{H2_column_density}), for which 
the column densities are very well constrained thanks to the damping wings.We detect rotational levels up to $J=6$ (and possibly $J=7$), the spectral resolution does not allow us to obtain trustable column densities beyond $J=4$.  
In particular, the saturated nature of many H$_2$ lines together with the medium spectral resolution of X-shooter does not allow us to 
consider independent Doppler parameters for each rotational level when a possible increase of $b$ with $J$ has been shown to exist in different environments at low \citep{Lacour2005} and high redshift \citep{Noterdaeme2007}.  
Still the measurement of the first five rotational levels contains interesting information, which we discuss below. 

\begin{table}
\centering
\caption{Column densities of H$\rm _2$ in different rotational levels at $\zabs=2.4636$ towards J1513$+$0352 
\label{H2_column_density}}
\begin{tabular}{c c}
\hline \hline
Species        & $\rm \log N$ (cm$\rm ^{-2})$ \\
\hline
H$\rm _2$, J=0 & $\rm 20.97\pm0.02$ \\
H$\rm _2$, J=1 & $\rm 21.03\pm0.02$ \\
H$\rm _2$, J=2 & $\rm 19.25\pm0.04$ \\
H$\rm _2$, J=3 & $\rm 18.94\pm0.04$ \\
H$\rm _2$, J=4 & $\rm 18.20\pm0.09$ \\
H$_2$, total   & $21.31 \pm 0.01$\\
\hline
\end{tabular}
\end{table}

\subsection{Rotational levels of H$\rm _2$ }
\label{Rotational levels of H2}
\label{H2ext}
\begin{figure}[!ht]
\centering
   \includegraphics[width = 0.5\textwidth]{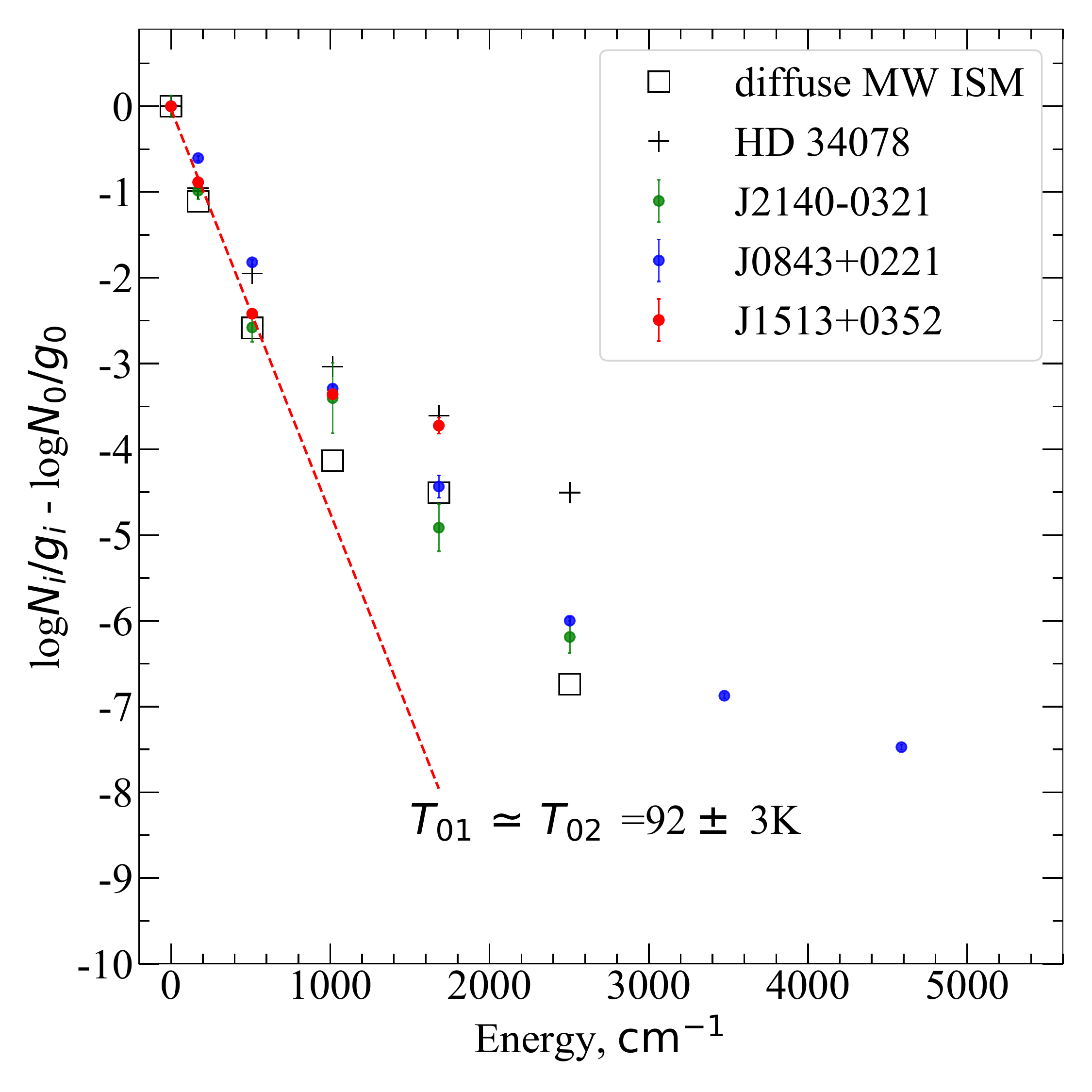}
     \caption{$\rm H_2$ excitation diagram of the diffuse molecular cloud towards J1513+0352 (red) compared with other environments (typical diffuse ISM in the MW shown as empty squares, the high-excitation case next to the bright star HD\,34078 shown with a plus sign; and two other ESDLAs, towards J0843+0221 (blue) and J2140-0321 (green)). 
     The red line shows the best fit to the J=0,1 and 2 rotational levels towards J1513+0352, indicating $\rm T_{\rm k} = 92 \pm 3$\,K.} 
     \label{Excitation_diagram}
\end{figure}

Figure~\ref{Excitation_diagram} presents the excitation diagram of $\rm H_2$, i.e. the relative population of H$_2$ rotational levels as a function of their energy with respect to the ground state. We find that both the $J=1$ and $J=2$ levels are consistent with a single excitation temperature, $T_{012} = 92\,\pm\,3$\,K. Since these low rotational levels are thermalised, their excitation temperature is known to be a good measurement of the kinetic temperature of the gas \citep[see][]{roy2006}, meaning that $T_k \approx T_{012}$. 
This temperature is only slightly higher than what is observed towards nearby stars ($T_{01} \sim 77$~K, \citealt{Savage1977}) but lower than what is typically seen in other high-$z$ H$\rm _2$-bearing DLAs \citep[$T\sim 150$~K;][]{srianand2005}, although there is also dispersion between the measurements, which may also reflect the dispersion in metallicity, with a trend of lower temperatures with higher metallicities. In addition, most previous measurements were derived from lower column-density systems and the lower temperature measured here is consistent with the trend of decreasing $T_k$ with increasing $N($H$_2)$ \citep[see e.g.][]{Muzahid2015, Balashev+17}. 

Higher rotational levels (i.e. $J\ge 3$) in turn have higher excitation temperatures. Such deviations from the Boltzman distribution is ubiquitous in the diffuse gas and generally explained by UV pumping and the fact that H$_2$ is produced in excited states at the surface of dust grains. 
A comparison between different diagrams in different environments is then useful to get at least a qualitative 
estimate of the UV field. 
In Fig.~\ref{Excitation_diagram}, the excitation of H$_2$ rotational levels towards J1513+0352 is compared with typical excitation seen 
in the Galactic disc \citep{Gry2002} and in a molecular cloud illuminated by a high UV field near a O9.5V star \citep{Boisse2005}. 
We also show the excitation diagrams for the high redshift systems towards J0843+0221 \citep{Balashev+17} and J2140-0321 \citep{Noterdaeme2015} that also have very large \HI\ column densities. The excitation of the DLA studied here is higher than the typical MW environment and is suggestive of a relatively strong UV field. An accurate measurement of the H$_2$ population in the high-rotational levels would be highly desirable, together with a measurement of the corresponding Doppler parameters. This should be possible with high resolution spectroscopy. Nevertheless, the high excitation inferred here suggests that 
the system towards J1513+0352 may represent gas in the star-forming disc of its associated galaxy. 

\subsection{CO content}

Carbon monoxide (CO) can only survive in cold H$_2$ gas. However, in spite of the very high H$_2$ column density observed 
here, CO lines are not detected. Using the procedure described in \citet{Noterdaeme2018}, we derive an upper limit 
of $\log N(CO) <13.60$~cm$^{-2}$ at the 3\,$\sigma$ confidence level. 
This translates to an upper limit on the CO/H$_2$ ratio of $\rm log(N({\rm CO})/N(\rm H_2))$ of $\rm < -7.8$, which is about two to three orders of magnitude lower than the typical value seen in the Milky Way at the same H$_2$ column density \citep[e.g.][]{Federman90, Sheffer08, Burgh09}.
As discussed by \citet{Balashev+17}, such an extremely low value is likely due to low metallicity of the gas. We indeed note that all CO detections to date at high-redshift have been obtained in high-metallicity gas ($Z\sim Z_{\rm \odot}$, \citealt{Noterdaeme2018}), although with low total hydrogen column densities. While high metallicities certainly favour the production of CO, through increased abundance of carbon and oxygen, but also an increased dust content, the low total column densities are in turn most likely reflecting a bias against high dust column densities that preclude the background quasars from being selected in optical surveys, as evidenced by several authors \citep[see e.g.][]{Fynbo2013, Krogager2016, Fynbo2017, Heintz2018}.
The low CO/H$_2$ ratio perhaps hints towards higher than Galactic cosmic-ray/X-ray ionization rates (see Fig. 16 and 17 of \citealt{Bialy2015}). However, a qualitative conclusion would require a detailed modeling that takes
  into account cloud geometry, covering factor, radiative trasfer models etc., which are beyond the scope of this paper.
The CO/\CI\, ratio is also low, with $\log N($CO$)/N(\CI) < -1.3$, which, in addition to the low metallicity, can also be explained by a high UV field to density ratio, as shown by \citet{Burgh09}. This is actually what we find from the physical conditions analyzed in Sect.~\ref{UV_Field_discussion}.

Interestingly enough, we remark that the system studied here is another evidence for the existence of "CO-dark" molecular gas at high redshift. Deriving the properties of a larger sample of such systems, together with determining a global 
census of strong H$_2$ systems would be highly desirable to provide an independent look at the molecular gas content of high-$z$ galaxies 
(see also \citealt{Balashev+18}). 

\subsection{Metallicity and depletion} 
\label{metallicity_and_depletion}

\begin{figure*}
\centering
   \includegraphics[width = \textwidth]{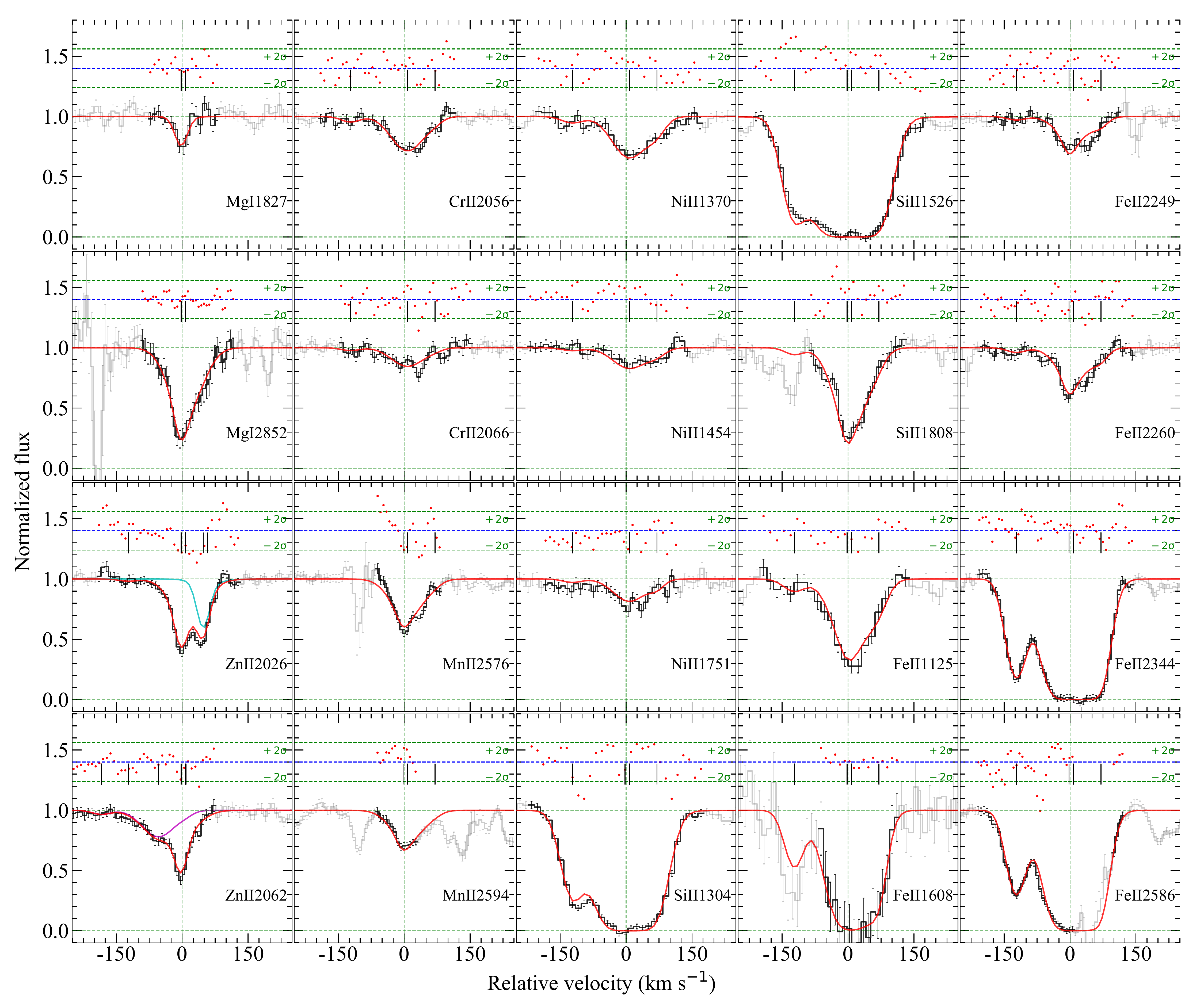}
     \caption{Absorption profiles of \ion{Si}{ii}, \ion{Ni}{ii}, \ion{Fe}{ii}, \ion{Zn}{ii}, \ion{Cr}{ii}, \ion{Mn}{ii} and \ion{Mg}{i}. The grey line represents the data with parts used to constrain the fit shown in black. The red line represents the synthetic profile. The black tick marks on top of the profile show the position of individual velocity components. The top box represents the residuals in the $\rm \pm 2\, \sigma$ range. The individual contributions from \ion{Mg}{i} and \ion{Cr}{ii} in the box showing \ion{Zn}{ii}($\lambda$=2026 and 2062) profiles are displayed in cyan and magenta respectively.
     }     
     \label{Metal-1}
\end{figure*}

We derive the metal abundances in the neutral gas from measuring the column densities of low ionization species. 
We fitted the absorption lines from \ion{Si}{ii}, \ion{Ni}{ii}, \ion{Fe}{ii}, \ion{Zn}{ii}, \ion{Cr}{ii} and \ion{Mn}{ii} assuming the same kinematic profile (i.e. we tied the redshifts and Doppler parameters, $b$, for each component). 
\ion{O}{i} and \ion{Mg}{ii} are also detected in the DLA, but the observed lines (specifically \ion{O}{i}~$\lambda1302$, \ion{Mg}{ii}~$\lambda2796$ and \ion{Mg}{ii}~$\lambda2803$) are saturated, so that we cannot get meaningful constraints on the corresponding 
column densities. We found that the metal absorption profiles are well reproduced using a four components model, one of which being narrow and corresponding to the \CI\ absorption. We also checked that using more components does not provide any improvement in the fit and results in the same total column densities within errors. 
In order to take into account the contribution of Mg\,{\sc i}~$\lambda$2026 to the Zn\,{\sc ii}~$\lambda$2026 profile, we also included Mg\,{\sc i} lines (\ion{Mg}{i}~$\lambda1747$, \ion{Mg}{i}~$\lambda1827$ and \ion{Mg}{i}~$\lambda2852$) while fitting the metals. 
The velocity profiles of the metal lines are shown in Fig.~\ref{Metal-1} together with the best-fit multicomponent Voigt profile. 
The total column densities, gas-phase abundances and abundance relative to zinc are summarized in Table ~\ref{total_column_density}.

\begin{table}
\centering
\caption{Summary of column densities and abundances \label{total_column_density}}
\begin{tabular}{c c c c} 
\hline \hline
Species        & $\rm \log N$ (cm$\rm ^{-2})$  & [X/H]\tablefootmark{a} & [X/Zn]\tablefootmark{b} \\
\hline
H\,{\sc i}     & $\rm 21.83\pm0.01$        &                        & \\
H$\rm _2$          & $\rm 21.31\pm0.01$        & $f$=$\rm 0.38\pm 0.02$\tablefootmark{c} &  \\
\HI+H$_2$ & $22.04 \pm 0.01$ & & \\
C\,{\sc i}     & $\rm 14.90\pm0.14$        &  & \\ 
Mg\,{\sc i}    & $\rm 14.16\pm0.12$        &  & \\ 
Zn\,{\sc ii}   & $\rm 13.83\pm0.23$        & $\rm -0.84\pm0.23$         & \\
Si\,{\sc ii}   & $\rm 16.92\pm0.24$        & $\rm -0.63\pm0.24$         & $\rm 0.21\pm0.33$\\
Cr\,{\sc ii}   & $\rm 13.78\pm0.02$        & $\rm -1.90\pm0.03$         & $\rm -1.07\pm0.23$\\
Mn\,{\sc ii}   & $\rm 13.21\pm0.03$        & $\rm -2.31\pm0.03$         & $\rm -1.47\pm0.23$\\
Fe\,{\sc ii}   & $\rm 15.46\pm0.03$        & $\rm -2.06\pm0.03$         & $\rm -1.22\pm0.23$ \\
Ni\,{\sc ii}   & $\rm 14.24\pm0.02$        & $\rm -2.02\pm0.03$         & $\rm -1.18\pm0.23$\\
\hline
\end{tabular}
\tablefoot{\tablefoottext{a}{Observed gas-phase abundances. Note that since the H$_2$ fraction is high, the total (\HI+H$_2$) hydrogen content is considered here.} 
\tablefoottext{b}{Depletion relative to Zn}
\tablefoottext{c}{{\sl Overall} molecular fraction through the system.}
}
\end{table}

\begin{figure}
   \includegraphics[width = 0.4\textwidth]{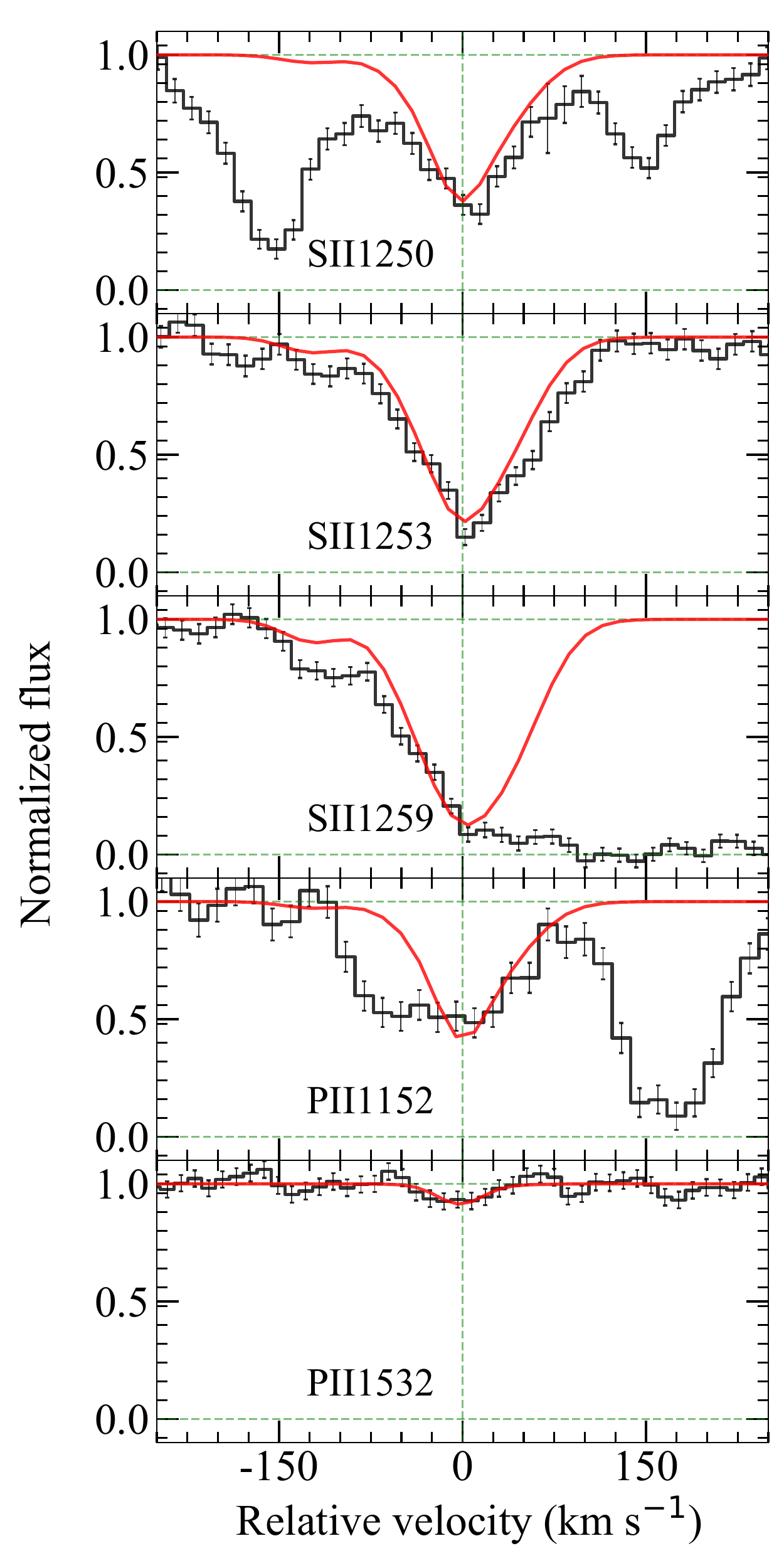}
\caption{Expected velocity profiles of \PII\ and \SII\ lines (red) overplotted on the observed spectrum (black). We emphasize that this is not a fit, but the scaling of the derived zinc profile assuming solar phosphorus-to-zinc and sulphur-to-zinc ratios. \label{PIISII}}
\end{figure}

We use the abundance of zinc, which is a non-refractory element, as an estimate for the DLA metallicity,  $\rm [Zn/H]$ = $-0.84 \pm 0.23$, where the relatively large uncertainty is due to the observations of the main narrow component at medium spectral resolution. 
  The abundance of silicon is apparently possibly higher than that of zinc albeit consistent within errors, 
which would be surprising since silicon tends to be mildly depleted. Indeed, \citealt{DeCia2016} showed that 
silicon should be depleted by a factor of two compared to zinc at the estimated metallicity. However, because the \SiII\ lines are intrinsically saturated and the narrow component hidden within a broad one, we caution its column density may be overestimated and hence, we prefer not to rely on \SiII. 
We also detect \PII\ and \SII\ in the system. However, we don't use these lines to constrain the fit as all \SII\ lines 
are located in the Lyman-$\alpha$ forest. The strongest \PII\ line is also located in the Lyman-$\alpha$ forest and another line which is outside the forest is close to the noise level. Still, these species provide a nice consistency check, in particular, since they are little or not depleted. We 
calculated the \SII\ and \PII\ profiles assuming a Solar abundance relative to zinc and found a very good  
match between the expected profile and the data, see Fig.~\ref{PIISII}. This further confirms our metallicity measurement. 
The derived metallicity is quite typical of DLAs at that redshift, although slightly higher than the cosmological mean metallicity value \citep{Rafelski2014, DeCia2017}. It is in turn significantly higher 
than the metallicity in the system towards J\,0843+0221 \citep{Balashev+17}, that has similar \HI\ and H$_2$ content. 

\subsection{Dust extinction of the background quasar's light }
\label{Dust_content}

The wide wavelength range and good flux calibration of the X-Shooter spectrum allows us to measure the reddening of the background quasar's light due to dust associated with the ESDLA. We use the template-matching technique as used in several studies of quasar absorbers \citep[see e.g.][]{Srianand200821cm, Noterdaeme2009co, Ma2015}. We allow for intrinsic quasar shape variation directly from fitting, as in \citet{Krogager2016}, instead of correcting the observed values using a control sample. More specifically, the observed quasar spectrum is matched by a quasar template, here taken from \citet{Selsing2016}, reddened using extinction laws applied at the DLA redshift. 
The extinction laws are described using the standard parametrisation by \citet{FitzpatrickandMassa2007} and the parameters obtained for different environments by \citet{Gordon2003}. 
We then fitted the model $T(\lambda)$ to the observed spectrum using 

\begin{equation}
T(\lambda)\,=\,f_0\times\,T_{\rm QSO} \Big( \dfrac{\lambda}{\lambda_0} \Big)^{\Delta\beta} \times \exp\Big[-0.92\,\epsilon(\lambda)\,E(B-V) \Big]
\end{equation}

\noindent where $T_{QSO}$ denotes the unreddened quasar template at the quasar redshift and 
$\epsilon(\lambda)$ is the wavelength-dependent reddening law. The parameters we vary are $E(B-V)$, the 
colour excess at the DLA redshift, $\Delta\beta$, the power-law slope relative to the intrinsic slope of the quasar template, and $f_0$, which is a constant scaling of the intrinsic quasar brightness.

\begin{figure*}
\centering
   \includegraphics[width = \textwidth]{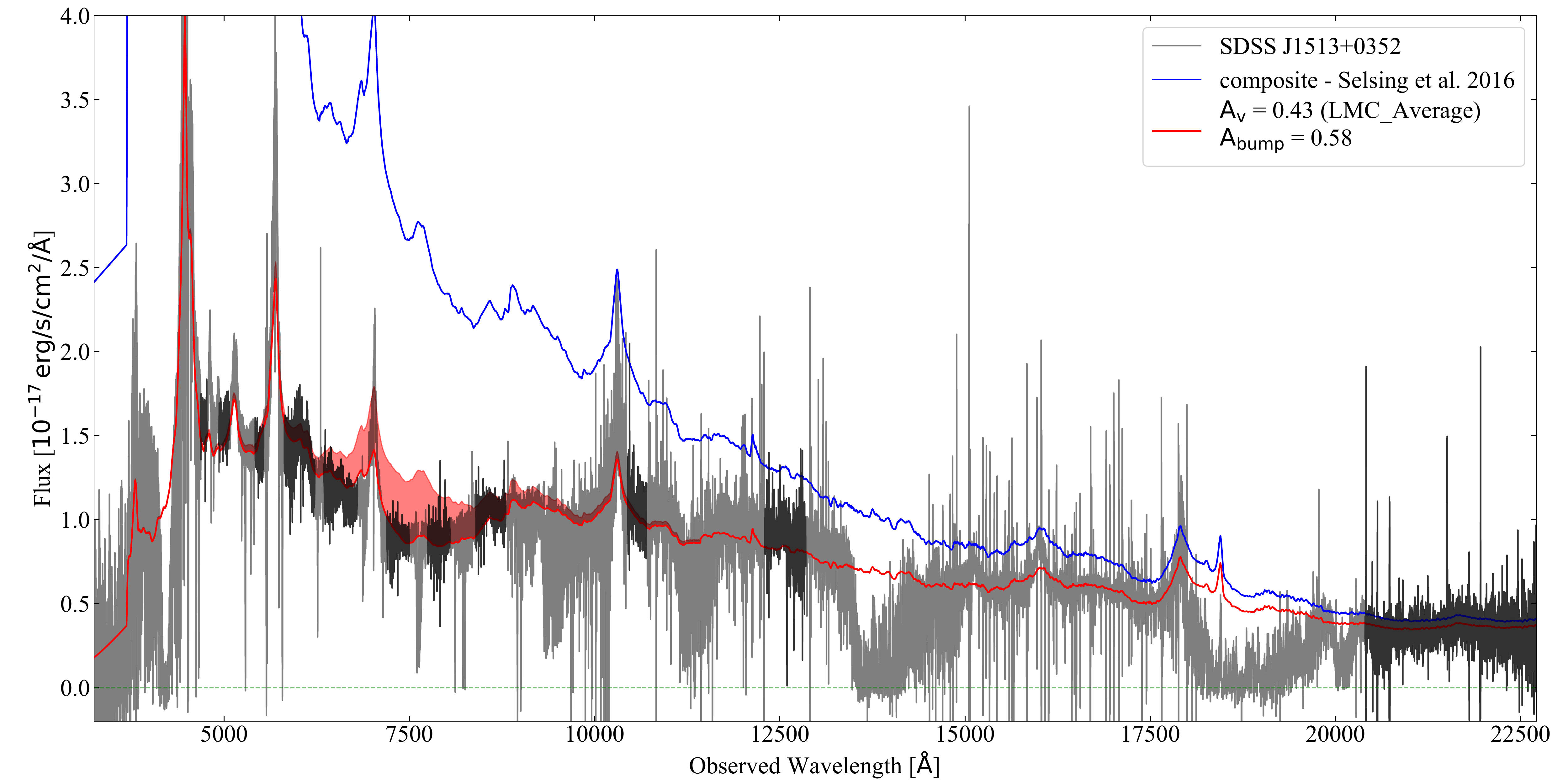}
   \caption{Measurement of the extinction of the background quasar by dust in the DLA. The X-Shooter spectrum is shown in grey with the continuum regions used to constrain the fit highlighted in black. The blue spectrum shows the scaled but unreddened quasar composite from \citet{Selsing2016}. The thick red line shows the composite reddened by a LMC average extinction law with $A_{\rm V}(DLA) = 0.43$. The shaded region represents the strength of the 2175 \AA\, bump ($\rm A_{bump}$(DLA)=0.58 $\rm \pm$ 0.003), that is the area between the reddened composite and the same reddened composite where the bump component (c$_3$) has been set to zero. \label{Dust_extinction}}
\end{figure*}

We found that the best-fit was obtained using the "LMC Average" extinction curve with an $A_{\rm V} \approx 0.43$, see Fig.~\ref{Dust_extinction}. The statistical 
error from the fit is only of 0.01 mag, but this does not represent the systematic error due to the spread of intrinsic quasar spectral slopes. A more realistic estimate of the uncertainty is of the order of 0.1~mag \citep[e.g.][]{Noterdaeme17}, consistent with the dispersion in intrinsic quasar spectral slope observed in the literature \citep{VandenBerk2001,Krawczyk2015}.

The absorption bump that exists in the LMC dust law is clearly visible at 2175~{\AA} in the DLA rest-frame. We quantified the strength of the bump following the prescription of \citet{Gordon2003} by following the parametrization from \citeauthor{FitzpatrickandMassa2007}, keeping $c_{3}$ and the other parameters fixed to the LMC value. The bump strength, which corresponds to the shaded area in Fig.~\ref{Dust_extinction} is then given by 

\begin{equation}
\rm A_{bump} = \Big(\dfrac{\pi\,c_{3}}{2\,\gamma\,R_V}\Big)\,A_V 
\end{equation}

\noindent where, $\rm c_{3}$ and $\rm \gamma$ correspond to the amplitude and width of the 2175 \AA\, bump. We obtained $A_{\rm bump} \simeq 0.58$, which is among the high values seen in other quasar absorption systems such as \CI\ systems \citep{Ledoux2015}, which are also known to have high metallicities \citep{Zou2018} and high molecular content \citep{Noterdaeme2018}. 
However, the high $A_{\rm V}$ and $A_{\rm bump}$ observed here are mostly due to the very large column density of the system. 

Indeed, the dust-to-gas ratio, $A_{\rm V} / N(H) \simeq 4\,\times\,10^{-23}$~mag\,cm$^2$ remains modest and similar to the average value of typical DLAs ($\sim 2-4\,\times\,10^{-23}$~mag\,cm$^2$, \citealt{Vladilo2008}) yet much smaller than 
the local ISM ($0.45\,\times\,10^{-21}$~mag\,cm$^2$, \citealt{Watson2011}) or \CI-selected high-metallicity DLA systems at high-$z$ ($\sim 10^{-21}$~mag\,cm$^2$, see \citealt{Ledoux2015} and \citealt{Zou2018}). 
We note that the dust-to-gas ratio, relative to the Milky-Way, is about 0.09 when the metallicity is about 0.14 Solar. While the uncertainty on {\bf our} measurement remains large, it may also indicate a non-linear relation between dust-to-gas ratio and metallicity, as 
seen by \citet{Remy-ruyer2013}.  

\section{Detection of the associated galaxy \label{galaxy}}
\subsection{Ly-$\alpha$ emission}

\begin{figure}
\centering
   \includegraphics[width = 0.45\textwidth]{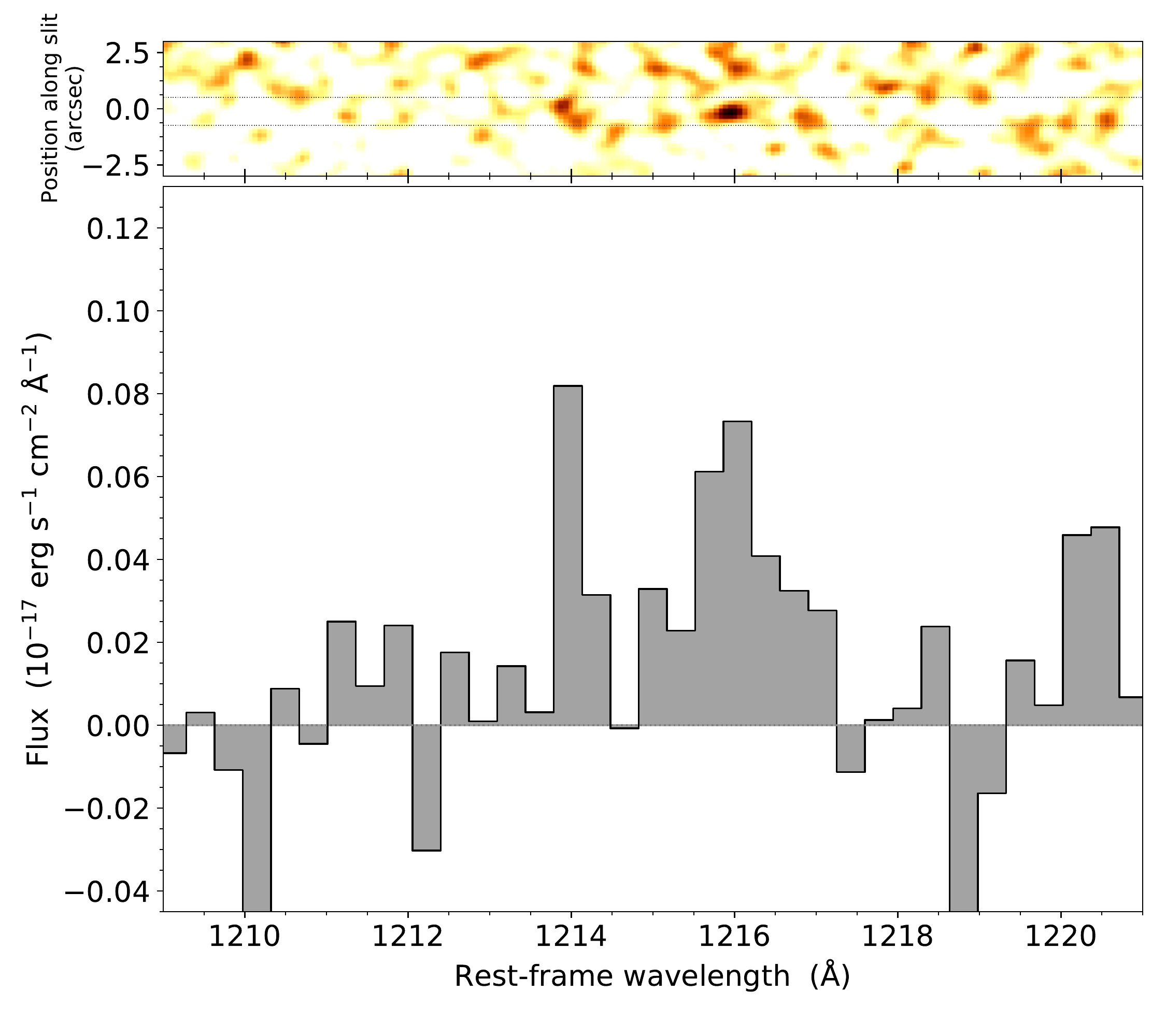}
     \caption{Ly $\rm \alpha$ emission in the DLA absorption trough shown in 1D (bottom, using a flux-conservative re-binning) and the 2D spectrum (top). The 2D data have been further smoothed by a Gaussian filter with $\sigma$ of 1.8 pixel. The extracted flux is $\rm \sim\,1.9\,\pm\,0.7\,\times\,10^{-18}\,erg\,s^{-1}cm^{-2}$.}
     \label{2-D_emission}
\end{figure}

\begin{figure}
\centering
   \includegraphics[width = 0.45\textwidth]{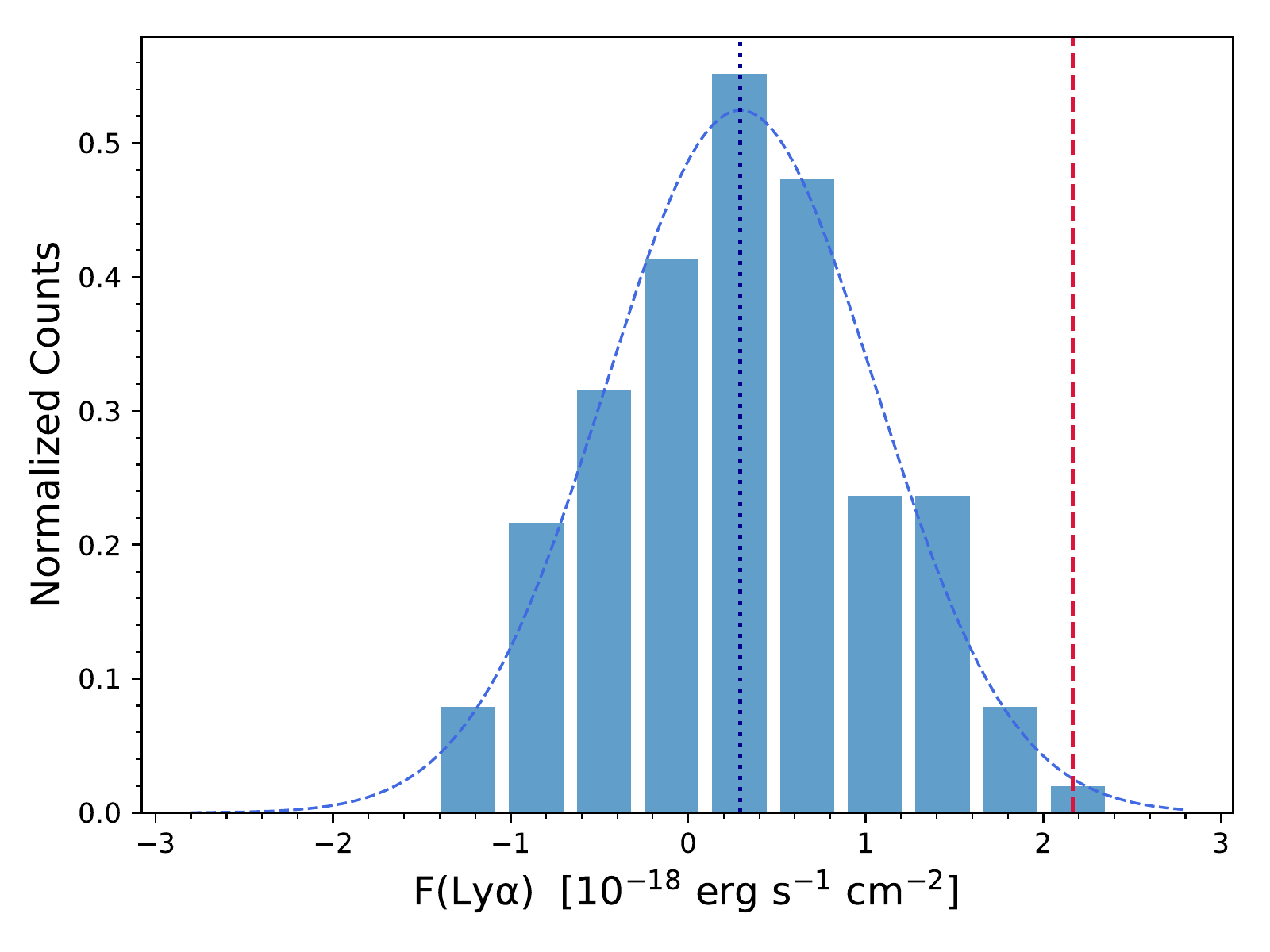}
     \caption{
     The histogram shows the distribution of fluxes measured in apertures randomly located around the detected emission. The apertures have the same size as that used for the Ly$\alpha$ measurement. The mean of the distribution (vertical dotted line) provides an estimate of the zero flux level and the dispersion provides a more conservative estimate of the noise in the aperture. The red, dashed, vertical line shows the measured value of Ly$\alpha$ emission associated with the DLA.}
     \label{zero_level_1D}
\end{figure}

A faint Lyman-$\alpha$ emission is detected at the bottom of the DLA trough that 
is already seen as a non-zero flux in the unbinned 1D data (see Fig.~\ref{HI_and_H2}). The same excess appears 
clearer in Fig.~\ref{2-D_emission}, where we rebinned the data by 6 pixels in a flux-conservative manner. The emission is also seen in the combined 2D data as a blob at around 1216~{\AA} in the DLA rest-frame and a possible peak at 1214~{\AA}, both well aligned with the quasar trace (top panel of Fig.~\ref{2-D_emission}). We note here that all three observed position angles used for the observations differ by no more than $\pm\,10\rm ^{\circ}$, allowing us to use the combined 2D spectrum for the measurement.  
From this 2D spectrum, we measured a flux of $\rm 2.2\pm0.5\,\times10^{-18}\,erg\,s^{-1}cm^{-2}$ in an aperture of $\rm \pm400\,km\,s^{-1}$ in the wavelength direction and  $\rm \pm0.8$\,arcsec in the spatial direction relative to the centre of emission (corresponding to about twice the FWHM of the trace). Figure~\ref{zero_level_1D} shows the distribution of fluxes in apertures of the same size, randomly located in the fully absorbed part of the DLA trough (excluding the region near the emission) on either side of the quasar trace. In the spatial direction, the apertures extended roughly to about $\rm \pm2.5$~arcsec. We find that the zero flux level is slightly offset by $\rm 0.3\times10^{-18}\,erg\,s^{-1}cm^{-2}$, which we subtract from the measured Ly$\alpha$ flux.
The rms dispersion of the random aperture flux distribution additionally gives a more appropriate estimate on the Ly$\alpha$ flux uncertainty of $\sigma(F_{Ly\alpha}) = 0.7\times10^{-18}$\,erg\,s$^{-1}$cm$^{-2}$. Therefore, the corrected estimate of the observed Ly-$\alpha$ flux is
$F_{Ly\alpha}\, =\, 1.9 \pm 0.7\,\times\,10^{-18}$\,erg\,s$^{-1}$cm$^{-2}$ and the detection is significant at the 2.7\,$\sigma$ level.

Considering the standard relation between H$\alpha$ and star formation rate from \citet{Kennicutt1998}, 
case-B recombination theory \citep{Brocklehurst1971} for the conversion of Ly$\alpha$ flux into H$\alpha$ 
and introducing the escape fraction of \lya\ photons, $f_{esc}$, we obtain a constraint on the star formation rate:

\begin{equation}
\label{SFRly}
\rm SFR(Ly\alpha)\, (M_{\odot}yr^{-1})\,= \dfrac{0.908\times10^{-42}\,F(Ly\alpha)\times\,(4\pi\,d_{L}^{2})}{f_{esc}} 
\end{equation}

\noindent where, $\rm d_{L}$= 6.3 $\rm \times$ 10$^{28}$\,cm is the luminosity distance for $z=2.46$. 
For $f_{esc}=1$, Eq.~\ref{SFRly} provides a lower limit on the star formation rate
of SFR~$> 0.09$~M$_{\odot}$\,yr$^{-1}$.

\subsection{Impact parameter} 
\label{impact_parameter}

\begin{figure}
\centering
   \includegraphics[width = 0.45\textwidth]{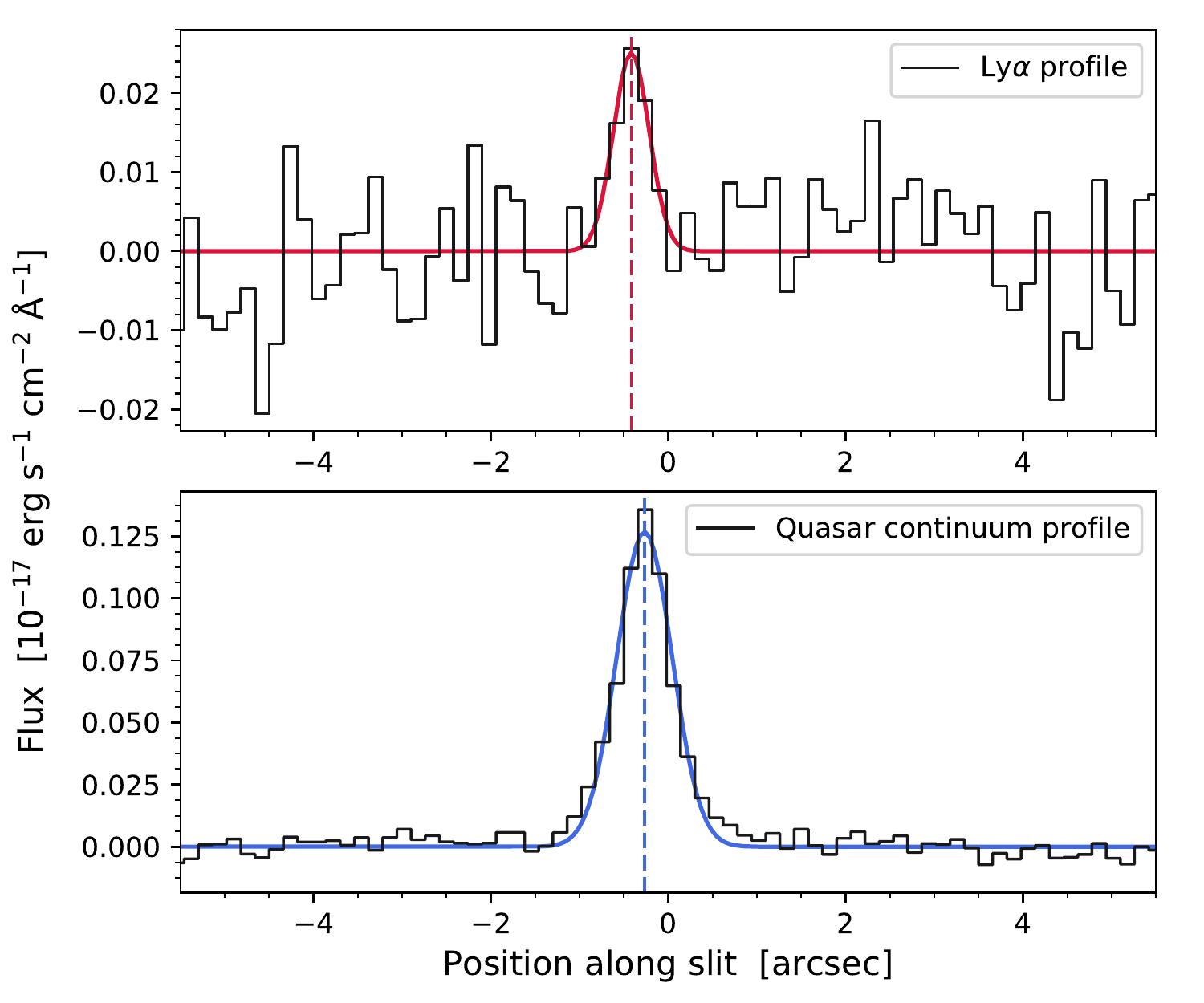}
     \caption{Spatial profile of the \lya\ emission (top) and quasar continuum (bottom) along the slit.
     The \lya\ profile has been averaged over 10 pixels centred at the \lya\ position (4211.8~\AA).
     The quasar continuum profile has been averaged over the wavelength range free of strong
     absorption ($\rm 4135-4150$~\AA).
     In each panel, we show the best-fit Gaussian model to the data. The vertical, dashed lines
     mark the centres of the profiles.
     }
     \label{fig:impact_parameter}
\end{figure}

\begin{figure}
\centering
\includegraphics[width=0.95\hsize]{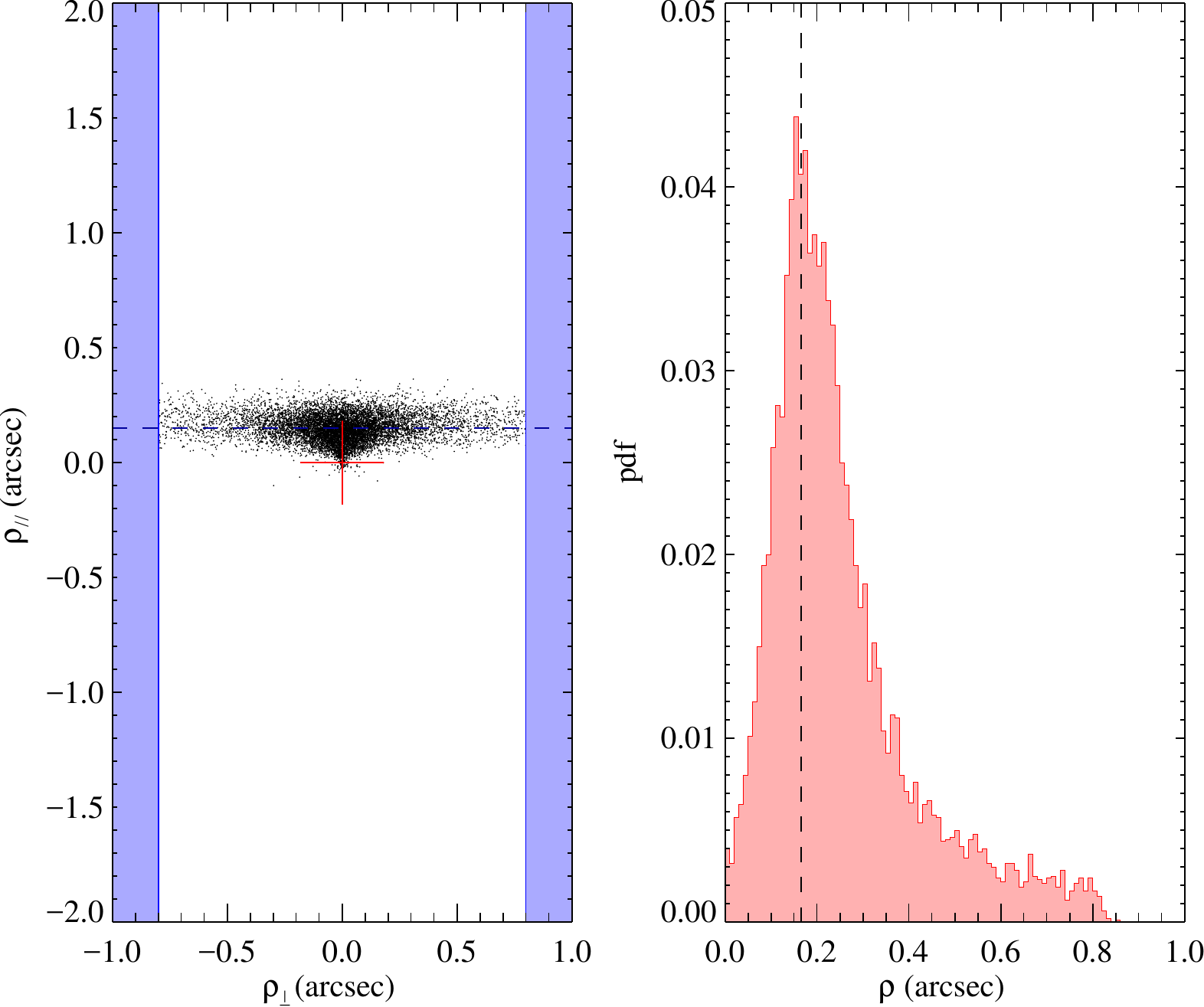}
\caption{Left: Layout of the 1.6$\arcsec$-wide slit (white region) with the position of the quasar marked as red cross. The dashed line represent our measured constraint on $\rho_{\parallel}$. $\rho_{\perp}$ is constrained to be 
within the slit width. The black points show the distribution of possible locations of the emission, taking 
into account a normally-distributed uncertainty on $\rho_{\parallel}$ and an uniformly-distributed random 
angle between quasar-galaxy and slit directions. Right: the probability distribution function of the impact parameter. \label{rhopdf}}
\end{figure}

While the similar position angles of each exposure allowed us to use the combined 2D spectrum to detect the emission, it is not 
possible to triangulate the emission as done when position angles differ by a larger amount \citep[see e.g.][]{Moller2004,Fynbo2010,Noterdaeme2012, Srianand2016, Krogager+17}. We can still estimate the impact parameter along the direction of the slit, $\rm \rho_{\parallel}$ from the spatial offset between the centroid of the Lyman-$\rm \alpha$ emission and the location of the quasar trace. We fitted Gaussian functions to the Ly-$\rm \alpha$ and quasar spatial profiles (collapsed along the wavelength dimension in the Ly-$\rm \alpha$ region and quasar continuum near to the DLA, respectively) and obtained $\rm \rho_{\parallel} = 0.15 \pm 0.06$\,arcsec, which corresponds to $\rm 1.24 \pm 0.5$\,kpc at the DLA redshift, see Fig.~\ref{fig:impact_parameter}.  Since the \lya\ emission is detected, we can further constrain its impact parameter in the direction perpendicular to the slit, $\rm \rho_{\perp}$, to be less than half the slit-width, i.e. $\rm \rho_{\perp} < 0.8$ arcsec,  or equivalently $\rho_{\perp} < 6.6~kpc$ in physical projected distance at the DLA redshift. This is a strict upper-limit since there is no particular reason for the impact parameter perpendicular to the slit being much larger than that 
along the slit length since the position angle were chosen without any prior. Indeed, assuming a uniformly-distributed random angle of the quasar to \lya\ emission direction with respect to the direction of the slit we obtain the most probable value of the impact parameter to be $\rm \rho \simeq 0.17\,\pm\,0.11$\,arcsec, or 1.4 $\rm \pm$ 0.9 kpc (see Fig.~\ref{rhopdf}). In the following, we will assume this to be the true impact parameter of the galaxy. We caution that it is still possible that the \lya\ emission arises only from the outskirts of a galaxy that would be located mostly outside of the slit, although the probability distribution function shown on Fig.~\ref{rhopdf} suggests this is highly unlikely. 

\subsection{NIR emission}
Since the NIR arm of X-shooter covers the expected position of the redshifted oxygen and Balmer emission lines, we performed a subtraction of the spectral trace of the quasar to search for [\ion{O}{ii}], [\ion{O}{iii}] doublet, H$\rm \beta$ and H$\rm \alpha$. We do not detect any emission line, but we are able to put 3-$\rm \sigma$ upper limits on the H$\alpha$ and [\ion{O}{ii}] fluxes of $\rm < 2.6\times10^{-17}\,erg\,s^{-1}cm^{-2}$ and $\rm < 1.3\times10^{-17}\,erg\,s^{-1}cm^{-2}$, respectively. The aperture used for the emission lines in the NIR is $\rm \pm\,200\,kms^{-1}$ along the wavelength axis and $\rm \pm\,0.6$ arcsec (corresponding to twice the FWHM of the trace) in the spatial direction.
Using \citet{Kennicutt1998}, we can convert these line fluxes into dust-uncorrected star formation rates  
$\rm SFR < 6.6$\,Myr$^{-1}$ and $\rm SFR < 5.9$\,Myr$^{-1}$ for H$\rm \alpha$ and [\ion{O}{ii}], respectively. These limits on the SFR are consistent with the lower limit inferred from Ly$\rm \alpha$ and it also places a lower limit on the escape fraction, $\rm f_{esc}$ > 1.5 \%.

\section{Discussion}
\label{Discussion}

\subsection{Does the line of sight pass through the main disc of the associated galaxy? }

\label{Proximity_of_high_N(HI)_gas}
There has been growing evidence that the column density observed along a given line of sight is anti-correlated with the impact parameter between this line of sight and the centroid of the associated galaxy. 
More specifically, low column densities are found over a wide range of impact parameters, while large column densities are 
restricted to the inner regions of galaxies (see Fig.~8 of \citealt{Krogager+17}). This is, not only, evident from 21-cm maps of atomic hydrogen in well-formed nearby galaxies \citep{zwaan2005, Braun2012}, but also in hydrodynamical simulations of gas at high redshift \citep[e.g.][]{Pontzen2008,Yajima2012,Altay+13,RahmatiandSchaye2014}
and direct detection of high-$z$ galaxies responsible for DLA absorption \citep{Krogager+17}. In 
addition, \citet{noterdaeme2014} have statistically shown that extremely strong DLAs, with column densities in excess 
of $N(\HI)> 5\times 10^{21}$~\cmsq\ are located on average at impact parameters less than 2.5~kpc from Lyman-$\alpha$ emitters (although it may be much larger in a fraction of systems, \citep[see][]{Srianand2016}. Here, the small impact parameter and the large hydrogen column density fits well within this picture. Indeed, 
\citet{RahmatiandSchaye2014} predict a 1$\sigma$ range of 1 $\rm \lesssim\, \rho\, \lesssim$ 6\,kpc for the observed 
$\rm \log N(HI)=21.83$ and we observe a most probable value of $\rho \sim 1.4$~kpc. 
With such a small impact parameter, it is reasonable to think that the line of sight 
passes through the main disc of the associated galaxy\footnote{While galaxies are 
not necessarily yet well-formed at high-$z$, we here just refer as "disc" to the 
region where most stars are found.}.
\citet{Paulino-Afonso2017} estimated that the effective radius, $\rm r_{eff}$ of Lyman-$\rm \alpha$ 
emitting galaxies at the DLA redshift ($\rm z\,\sim\,2.5$) is $\rm r_{eff}\,\sim\,1$~kpc, which is consistent with our measured impact parameter since the effective radius is defined so that it covers half 
of the total stellar light emitted from the galaxy, which then drops down exponentially. This suggests that 
the line of sight still could pass through the light-emitting region of the associated galaxy. 
We can also get a constraint on the spatial extent of the \lya\ emission of our DLA galaxy from the 2D data. 
The unresolved emission suggests that the radius of the \lya\ emission is smaller than the 
physical projected size corresponding to the seeing, i.e., $r<2.85$~kpc, which remain consistent with our above picture. 
In conclusion, the typical properties of LAEs at high-$z$, the measured small impact parameter and the large hydrogen column density suggest that the line of sight passes through the main galactic disc. We caution again that we cannot exclude the possibility that the slit covers only part of the extended \lya\ emission from a galaxy that would be located outside the slit. However, this would require an ad-hoc configuration, putting the galaxy at a peculiar position so that the centroid of the covered Ly-$\alpha$ emission falls at the edge of the slit and almost perpendicularly to the slit direction. This is unlikely given the small value of $\rho_{\parallel}$ and the large slit width, as estimated in Fig.~\ref{rhopdf}. It still remains possible that the detected Ly$\alpha$ emission corresponds to a star-forming region disconnected from a large galaxy, in which case we can consider the star-forming region to be a galaxy on its own (possibly part of a group) and our discussion, then, still stands. Considering the impact parameter between the quasar and the observed \lya\ emission as a lower limit to the galaxy extent and the unresolved 2D emission as an upper-limit, we then constrain the radius of the galaxy to be in the range ($\rm 1.4\, kpc\, < r <\, 2.85$\,kpc).

\subsection{On the relation between gas surface density and star formation rate }
\label{absorption_to_emission_relation}
\label{Kennicutt-Schmidt Law}

In this section, we discuss how our observations compares with the relation between the surface densities of gas and star formation rate as observed in the nearby Universe, i.e. the Kennicutt-Schmidt relation.
This is valid over galactic scales and expressed as $\Sigma_{SFR} \propto \Sigma_{gas}^{n}$ with $n\sim 1.4$
\citep{KennicuttandEvans2012}. Here, we don't have access to the disc-averaged gas surface density, but a probe along a single line of sight so that we must assume it is representative of the mean column density over the galaxy extent. The total (\HI+H$_2$) column density we observe then corresponds to $\Sigma_{gas} \simeq 86$~M$_{\odot}$\,pc$^{-2}$.

From the detection of the Lyman-$\rm \alpha$ emission and the 3-$\rm \sigma$ upper limit from the non-detection of [O\,{\sc ii}], we obtained a conservative range on the star formation rate to be $0.1 < SFR < 5.9$~M$\rm _{\odot}\,yr^{-1}$. This range actually corresponds to an escape fraction ranging from 100\% (if SFR=0.1 M$\rm _{\odot}\,yr^{-1}$) to 1.5\% (if SFR=5.9 M$\rm _{\odot}\,yr^{-1}$). However, the typical escape fraction for the overall population of emission-selected galaxies at this redshift has been estimated to be more around 5\% \citep{Hayes2011} but can be much larger in low mass \citep{Fernandez2011} and compact \citep{Paulino-Afonso2017} galaxies. Indeed, for typical Lyman-$\rm \alpha$ emitters, the escape fraction is found to be about $\rm f_{ esc}\,\sim 30\%$  (\citealt{Finkelstein2011}, \citealt{Nakajima2012}, \citealt{Nestor2013}) . A more realistic range for the star formation rate can then be estimated considering $f_{\rm esc} \sim 5- 30\%$, from which we get $\rm 0.2<SFR<2.4$~M$_{\odot}$\,yr$^{-1}$. Combining these constraints with those on the galaxy extent, this translates to a surface density of star formation rate in the range $\rm -2.1<\log \Sigma_{SFR} (M_{\odot}\,yr^{-1}\,kpc^{-2})<-0.4$, assuming disc-like geometry and ignoring projection effects (i.e. taking the galaxy's surface to be $\pi r^2$). The estimated surface star formation rate and the observe column density are therefore consistent with the observations in the nearby Universe \citep[see e.g. figure 11 of][]{KennicuttandEvans2012}.

Narrowing down the constraint on the surface star formation rate through deep observations at high spatial resolution together with assembling a large sample of such high column density systems may provide very unique information on the relations between total gas and star formation in normal galaxies in the distant Universe, when current observations mostly consider only the molecular reservoirs (estimated indirectly through CO emission) and in emission-selected galaxies.

\subsection{The \HI-$\rm H_2$ transition}
\label{Bialys_formalism}

A key process in the conversion of gas into stars in galaxies is the transition from the atomic to the molecular phase, i.e. the \HI-H$_2$ transition. While it is not absolutely clear whether the change in chemical state of the gas (atomic to molecular) is required for stars to form or if this transition is simply the consequence of a change in physical state of the gas that also leads to star formation, it remains one of the most fascinating processes that occur in the ISM and so far has not been studied in depth at high redshift, despite important recent progress \citep{Noterdaeme17,Balashev+17,Balashev+18}. The transition from \HI-to-H$_2$ is however 
well studied theoretically and understood as a balance between the formation process of H$_2$ onto dust grains, and its destruction by UV photons. Detailed microphysics models link the column density above which the gas in a cloud turns molecular mostly with the metallicity, which acts as a proxy for the amount of dust in the gas and assuming a priori conditions of a cold neutral medium \citep[see e.g.][]{Krumholz2009, Sternberg2014, Bialy2016}. 
More recently, \citet{Bialy2017} have related the total atomic column density in the cloud external layer to the density in the cloud, the average absorption 
cross-section of dust grains and the intensity of the UV field, based on the theoretical work by \citet{Sternberg2014}. 

The \ion{H}{i} surface mass density ($\rm \Sigma_{\ion{H}{i}}$) of an optically thick, uniformly dense two sided slab irradiated by a far-UV flux is given as:

\begin{equation}
\label{Bialy1}
\rm \Sigma_{\ion{H}{i}} = \dfrac{6.71}{\tilde{\sigma}_{g}} 	ln\, \left( \dfrac{\alpha G}{3.2} + 1 \right) M_{\odot} pc^{-2},
\end{equation}

\noindent where $\rm \tilde{\sigma}_{g}\,\equiv\,\sigma_{g}/(1.9\,\times\,10^{-21}\,cm^{2})$ is the dust grain Lyman-Werner (LW = 11.2 - 13.6 eV, 911.6 \AA - 1107 \AA) photon absorption cross section per hydrogen nucleon normalised to the fiducial Galactic value, $\rm \alpha$ is the ratio of the unshielded $\rm H_2$ dissociation rate to $\rm H_2$ formation rate, and G is an average $\rm H_2$ self-shielding factor in dusty clouds (see \citealt{Bialy2016} for details on each parameter). The product $\rm \alpha G$ is expressed as: 

\begin{equation}
\label{Bialy5}
\alpha G\, =\, 0.59\,I_{UV}\,\Bigg( \dfrac{100\,{\rm cm^{-3}}}{n_{\rm H}} \Bigg)\,\Bigg( \dfrac{9.9}{1+8.9\tilde{\sigma}_{g}} \Bigg)^{0.37} 
\end{equation}

\noindent where $n_{\rm H}$ is the cloud hydrogen density and $\rm I_{UV}$ is the intensity of radiation field expressed in units of the Draine field \citep{Habing1968}.

This theoretical model has been successfully used to interpret the observations of molecular clouds in our galaxy \citep{Bialy2015A} but not yet to absorption-line systems in the 
distant Universe. Doing so is of particular interest since the available observables are very different locally in emission and at high redshift in absorption. For example, in the study of the molecular cloud complex W43 by \citet{Bialy2017}, the \HI\ column density has been obtained from 21-cm emission map by \citet{Bihr2015}, corrected for continuum emission and optical depth effects; the UV field in W43 is estimated from dust emission maps assuming thermal equilibrium of dust grains, the density is estimated from typical values predicted for  pressure equilibrium between cold neutral medium and warm neutral medium or from taking the averaged observed \HI\ surface density and dividing by an estimate of the radius of the molecular complex.  Finally, a fiducial value is taken for the dust-grain absorption cross section. 
Here, we discuss the derivation of the three parameters $\rm \tilde{\sigma_g}$, $\rm I_{UV}$ and $n_{\rm H}$ entering Eq.~\ref{Bialy1} using our observables measured in Sect.~\ref{analysis}.

\subsubsection*{Dust-to-gas ratio, $\rm \tilde{\sigma}_{g}$ }

The Lyman-Werner dust absorption cross section per hydrogen nucleon, relative to the Galactic value can be expressed as:

\begin{equation}
\label{Bialy4}
\tilde{\sigma}_{g}\, =\, 4.8\,\times\,10^{20}\,\dfrac{A_{\rm LW}}{N({\rm H})}\,\rm cm^2
\end{equation}

\noindent where $N$(H) (= $N(\HI) + 2N($H$\rm _2$)) is the total hydrogen column density, $A_{\rm LW}$ is the extinction in the Lyman and Werner band, which is the wavelength range of interest for the shielding of H$_2$. We derive $\rm A_{LW}$ at the centre of the band ($\rm \lambda_{LW}$(rest)=1010~\AA, $\lambda_{LW}$(DLA)=3496 \AA) using $A_{\rm V} = 0.43$ (from dust reddening, Sect.~\ref{Dust_content}) using the best-fit extinction law, which we found to be that of the Large Magellanic Cloud. We obtain $A_{\rm LW} = 2.1$ and hence $\tilde{\sigma}_{g} \simeq 0.1$ 
with  a statistical uncertainty of about 25\%. 
However, to take into account possible systematics intrinsic to the derivation of $A_{\rm V}$, we will consider a factor of 2 range on $\tilde{\sigma}_{g}$. We finally note that this value is also consistent with the expected approximate scaling with metallicity.

\subsubsection*{Hydrogen density ($n_{\rm H}$) and UV field density ($I_{\rm UV}$)}
\label{UV_Field_discussion}

We use measured excitation of fine-structure levels of \CI\ to estimate the physical conditions in the medium. The fine-structure levels of \CI\ can be excited by collisions, UV pumping and direct excitation by photons from the cosmic microwave background (CMB) \citep[see][]{Silva-and-Viegas2002}. The CMB radiation is fixed at any redshift following the adiabatic expression $T_{\rm CMB}(z)=(1+z)T_{\rm CMB}(z=0)$, as predicted by standard Big-Bang theory and constrained by observations \citep[e.g.][]{Noterdaeme2011}. The UV radiation field corresponds to the metagalactic background (taken from \citealt{Khaire2018}) plus the additional UV field, which corresponds to the galaxy. 
Collision rates depend on the number density of the medium and the temperature which we fixed to be $T = 92$\,K, based on the excitation of lower rotational levels of H$_2$. Therefore, we are left with two parameters, $n_{\rm H}$ and $I_{\rm UV}$ that are constrained simultaneously from the observed population ratio of \CI\ fine structure levels. We also assume a {\sl local} molecular fraction $f=1$, which is expected at such high H$_2$ column density. 

In any case, using the {\sl overall} molecular fraction $\avg{f}=0.38$ does not change the result significantly and would actually make little sense since this molecular fraction implies a perfectly uniform mixing of atomic and molecular hydrogen, i.e. no transition. 

\begin{figure}
\includegraphics[width=\hsize]{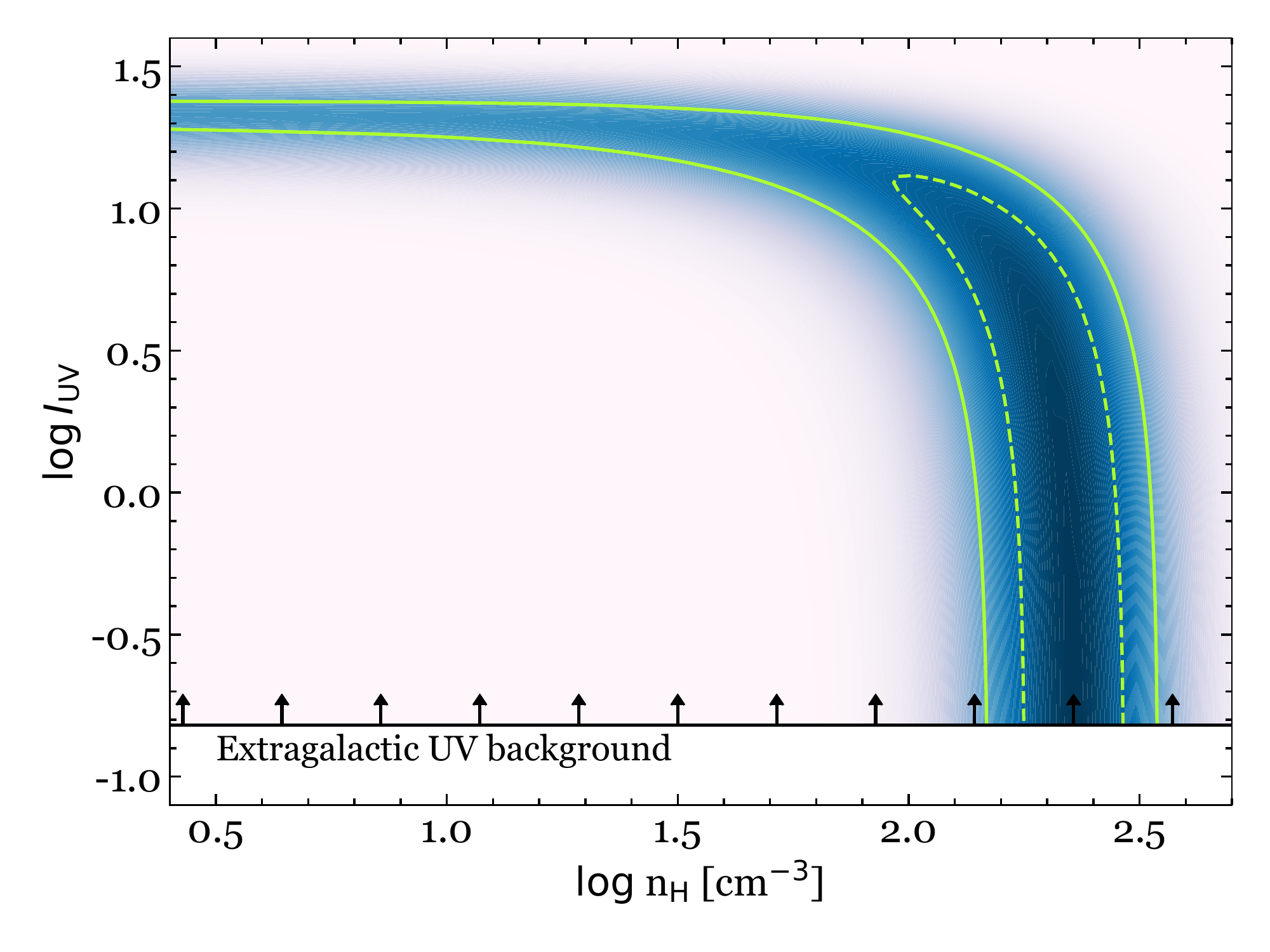}
\caption{Constraints on the hydrogen density--UV field plane from the excitation of \CI\ fine-structure levels. The probability density is represented by the colour intensity, with the 68\% and 30\% confidence level contours shown as solid and dashed green lines, respectively. The extragalactic UV background sets a lower limit to 
$I_{\rm UV}$ as shown by the horizontal black line with upward arrows. Note that attenuation effects are not taken into account here, and that the actual incident UV field can be a few times higher. 
     \label{popratio}}
\end{figure}

We calculated the relative ratio of \CI\ fine-structure levels using our own Python code, described in \citet{Balashev+17}, and similar to the {\sc PopRatio} code from \citet{Silva-and-Viegas2002}. We varied the density and UV field intensity over a 100$\times$100 grid
  with $n_{\rm H}$ and  $I_{\rm UV}$ ranging on a log-linear scale from respectively 1 to 1000~cm$^{-3}$ and 0.1 to 100~G$_{\odot}$, where G$_{\odot}$ represents the average UV radiation field of \citet{draine1978}. We then compared the calculated values with the observed ones.
Assuming the uncertainty on the observed $\log N(\CI*)/N(\CI)$ and $\log N(\CI**)/N(\CI)$ are Gaussian, we obtain the probability distribution in the ($n_{\rm H}$-$I_{\rm UV}$)-plane shown on Fig.~\ref{popratio}.

The \CI\ population ratio indicates a high hydrogen density, of the order of 250~cm$^{-3}$, in which case the excitation is dominated by collision and becomes insensitive to the UV field for a wide range of values. However, a lower hydrogen density is also allowed provided the UV field is high, around 20 times the Draine field, but this is less probable, as can be seen from the contours in Fig.~\ref{popratio}. We note here that this remains a simple diagnostic and that the calculation does not include radiative transfer. In other words, the estimated UV flux corresponds here to an unattenuated value due to shielding by dust and does not include effects of line saturation. Since \CI\ is found only in the cloud interior, with some amount of shielding, the $I_{\rm UV}$ values derived from the excitation of this species should be considered as a lower-limit to the actual UV field incident to the cloud, which could be up to a factor of a few higher, considering the measured dust-extinction (see Fig.~\ref{Dust_extinction}). 

We finally note that while the effect of cosmic rays is not taken into account here nor in the used transition theory, it is possible that cosmic rays maintain the molecular fraction to a value less than unity, even deeper than the \HI-H$\rm _2$ transition layer \citep[e.g.][]{Noterdaeme17}. Assuming only cosmic rays are responsible for the destruction of H$\rm _2$, the equilibrium between formation and destruction of H$\rm _2$ gives $N(\HI)/N($H$\rm _2) = \zeta(H\rm _2)/(R n) \simeq 3.3$. Then, assuming a formation rate on dust grains $R=3\times10^{-18}$ (a factor of roughly 10 smaller than in the Milky-Way), and for density obtained (at I$\rm _{UV}=0$) we get and upper limit on the H$\rm _2$ ionisation rate by cosmic rays to be about $\zeta(H\rm _2)=2.5\times10^{-15}$~s$^{-1}$. 

An independent and alternative way to estimate the incident radiation field is to scale it to the surface star formation rate, assuming that the field is uniform and isotropic: 

\begin{equation}
\rm \avg{I_{UV}} = \dfrac{\Sigma_{SFR}}{\Sigma_{SFR}(MW)} 
\end{equation}

\noindent Using the observed $\rm \Sigma_{SFR}$ in the DLA in the range 0.01 to 0.4 $\rm M_{\odot} yr^{-1}\,kpc^{-2}$ and the average Milky-Way value of $\Sigma_{SFR}$(MW) $\rm \sim\,0.003\,M_{\odot} yr^{-1}\,kpc^{-2}$ \citep{KennicuttandEvans2012}, the intensity of the mean radiation field in the DLA galaxy should be of the order of $I_{\rm UV} \sim 2.5$ to 100. 
We emphasize that this range is loose due to the uncertainties on the galaxy's size and star formation rate and that it represents only an average intensity field in the DLA galaxy, meaning that the local field incident to the cloud can actually be different. On the other hand, the H$_2$ high rotational levels actually also suggest a UV field higher than the average MW value.

\subsubsection*{Comparing observations to theory}

\begin{figure}
\centering
    \includegraphics[width = 0.5\textwidth]{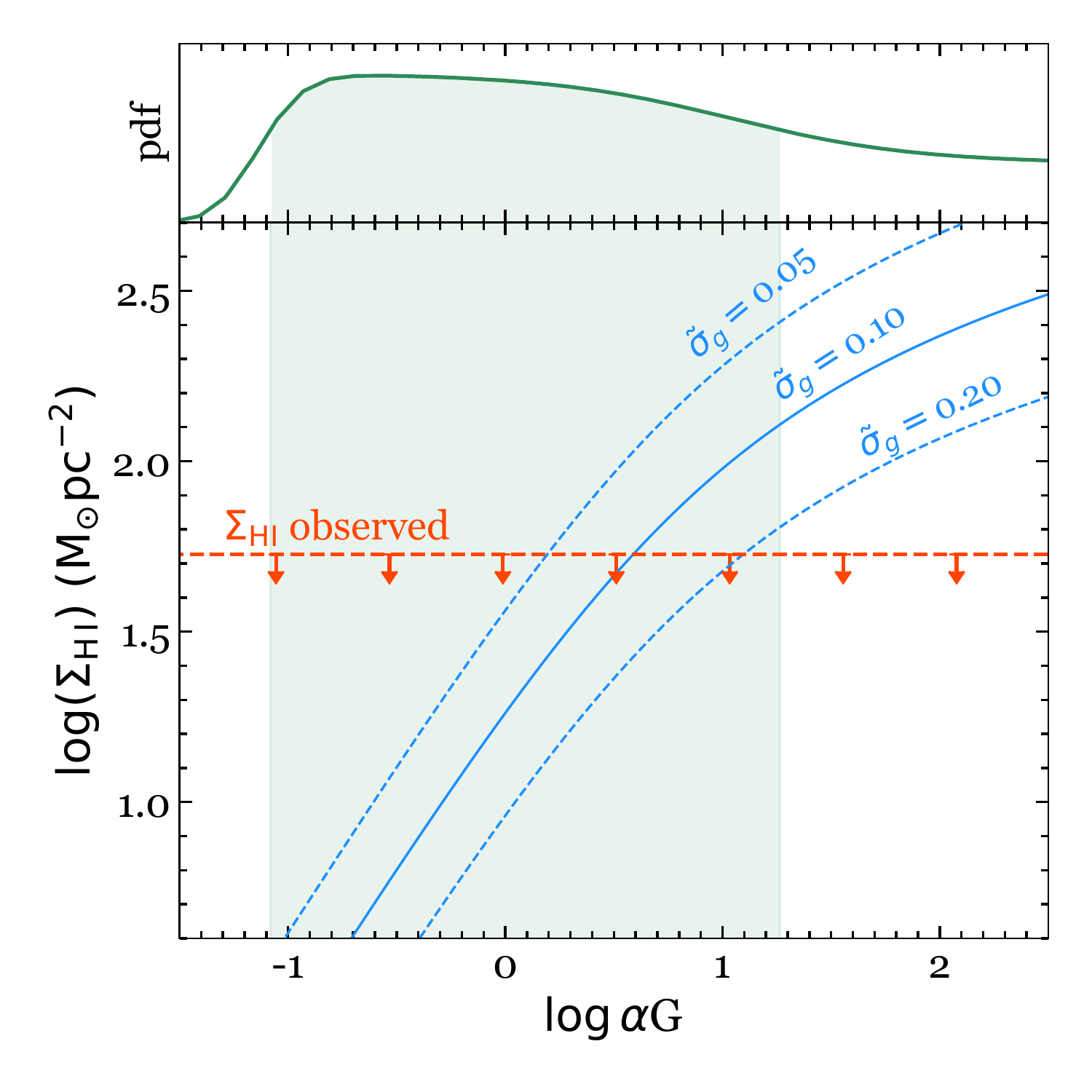}
     \caption{Comparison of \HI-H$_2$ transition theory predictions with observations. {\sl Top panel:} derived probability distribution function of $\rm \alpha G$ from excitation of \CI\ fine-structure levels. The green shaded region shows the interval estimate on $\rm \alpha G$ corresponding to 0.683 confidence level. {\sl Bottom panel:} theoretical \ion{H}{i} surface density of a single \ion{H}{i}-to-H$\rm _2$ transition layer as a function of $\rm \alpha G$ (Eq.~\ref{Bialy1}) for $\rm \tilde{\sigma}_{g}=0.1$ shown by solid blue line. The two dashed blue lines correspond to the derived range for $\rm \tilde{\sigma}_{g}$.       
     The observed {\sl total} \ion{H}{i} column density is shown as a red horizontal line with downwards arrows, since part of the \HI\ is likely to not be associated with the molecular cloud.}
     \label{sigmaHI}
\end{figure}

By propagating our constraints in the UV field -- density plane as shown in Fig.~\ref{sigmaHI}, we 
obtain the probability distribution function of $\alpha G$, as shown in the top panel of 
Fig.~\ref{sigmaHI}. We note that the dependence of $\alpha G$ on the dust is very small so that we simply used the value from the best-fit extinction here, i.e. $\tilde{\sigma_g} = 0.1$
The allowed $\alpha G$ range remains quite wide around a central value of about unity. 

The bottom panel of Fig.~\ref{sigmaHI} shows the theoretical dependence of the atomic hydrogen 
surface density as a function of $\alpha G$. Here, the dependence on the dust-to-gas ratio is linear. We illustrate the theoretical expectations for a factor of two around $\tilde{\sigma_g} = 0.1$ as well. 
The horizontal line corresponds to the observed total \HI\ column density integrated in the system. 
This should actually be considered as an upper limit since it may well contain atomic gas that does 
not belong to the envelope of the molecular cloud. 
Indeed, several components were seen in the low-ionisation metal lines. Assuming uniform metallicity, 
then the atomic gas column density associated to the molecular component decreases to 85\% of the overall 
value. Remarkably, the observed column density correspond to a value of $\alpha G$ that falls in the middle of our allowed range. In turn, it appears that the  high end of our derived $\alpha G$ range is inconsistent with the theoretical expectations. Similarly, the low end of our allowed range on $\alpha G$ would indicate a very small \HI\ column density, when the observed value is very large. This would only be possible if almost all of the \HI\ is not associated to the H$_2$ component, and in which case the metallicity of the narrow component would become unrealistically high. 
This shows that we can actually use the theoretical $\Sigma_\HI$ to constrain the UV field to density ratio. For the observed dust grain abundance, Eq.~\ref{Bialy5} simplifies as $I_{\rm UV}/(n_{\rm H} / {\rm 100\,cm^{-2}}) \approx \alpha G$. Using \HI\ column density from the simple redistribution of \HI\ in the different components assuming uniform metallicity, we obtain $\alpha G \approx 2$.

This, together with the constraint on the ($I_{\rm UV}$,$n_H$) plane from \CI\ excitation implies that the density is about $n_{\rm H} \simeq 250$~cm$^{-3}$ and the UV field is several times 
higher than the Draine field and possibly even an order of magnitude higher when taking into account the fact that \CI\ excitation provides a constraint on the attenuated flux (see previous section).
This is also consistent with the observed star formation rate as well as with the high excitation of H$_2$.

\section{Conclusion \label{Conclusion}}

We have presented the simultaneous detection of atomic and molecular gas and star formation activity in 
an absorption-selected galaxy.  
This allows us to study the global relationship between the gas-phase and the galactic properties on one side, and to investigate the physical conditions and microphysics of the \HI-to-H$_2$ transition on the other side. In particular, we find that the measured extremely high column densities of hydrogen (for both atomic and molecular form, $\log N(\rm cm^{-2}) = 21.83\pm0.01$ and $\log N(\rm cm^{-2}) = 21.31\pm0.01$, respectively) are consistent with the very small impact parameter to the centroid of the galaxy ($\rho\sim 1.4$\,kpc) and that the observed constraints on the star formation rate are consistent with the local Kennicutt-Schmidt law, although we note that our constraints remain quite loose.
Using the abundances and excitation of atomic and molecular species together with the dust extinction, we derive the chemical and physical conditions in the molecular cloud. Based on the measured zinc column density, we measured the average metallicity of this ESDLA to be about 15\% of the Solar value. 
Thanks to the long wavelength range of the X-shooter spectrum we measure associated extinction to be $A_{\rm V} \approx 0.43$ (assuming LMC average extinction curve). Using the relative population of C\,{\sc i} fine-structure levels we estimate the number density in the molecular cloud to be $n_{\rm H} \approx 250$~cm$^{-3}$, although a solution with high UV field and low density is also allowed. Unfortunately, independent constraints on the mean UV field in the galaxy from the observed star formation rate are not very stringent.  
The observed \HI\ column density, is in turn, consistent with the prediction from \HI-H$_2$ transition theory for a narrower range of physical conditions. This 
provides a constraint on the ratio of UV radiation intensity to hydrogen density, $\alpha G \approx 2$. This constraint complements that from \CI\ excitation in the UV intensity--hydrogen density plane and implies that
the UV field is several times that of our Galaxy, consistent with the average UV field estimated from the observed star formation rate and qualitatively consistent with the high excitation of H$_2$ in the cloud. In other words, we show that the observed \HI\ column density together with a measurement of dust abundance are able to provide important constraints on the physical conditions in a molecular cloud by comparing with the \HI-H$_2$ transition theory.   

Obtaining high-resolution data is desirable to ascertain the column densities, in particular of the high rotational levels of H$_2$, get better constraints on the absorption-line kinematics and possibly detect 
excited levels of more species, like \SiII$^*$ and \OI$^*$, as seen in other ESDLAs \citep{Kulkarni12,Noterdaeme2015}.
Detailed modelling of the cloud would then certainly help to further understand the physical conditions. 
Obtaining tighter constraints on the star formation rate through deep observations of nebular emission lines as well as on the macroscopic properties of the galaxy (in particular its extent) would also be of great value to further link the local chemical and physical conditions of the molecular cloud to those of star formation. Deep, spatially-resolved observations of the \lya\ emission together with radiative transfer modelling should also provide very useful information about the gas kinematics.
Finally, our single system suggests that absorption and emission observations of a sample of extremely strong DLAs are a promising way to link star formation to the properties of the interstellar medium in normal, high-redshift galaxies.

\begin{acknowledgements} 
We thank the referee for useful comments and suggestions. 
The authors are very grateful to ESO and in particular to the Paranal observatory's staff for carrying out our observations in service mode. 
AR, PN, PPJ, NG and RS gratefully acknowledge the support of the Indo-French Centre for the Promotion of Advanced Research (Centre Franco-Indien pour la Promotion de la Recherche Avanc\'ee) under contract no. 5504-2. The research leading to these results has received funding from the French {\sl Agence Nationale de la Recherche} under grant no ANR-17-CE31-0011-01 (project "HIH2" - PI Noterdaeme). SAB thanks the Institut d'Astrophysique de Paris for hospitality and the Institut Lagrange de Paris for financial support during the time part of this work was done. SAB is supported by RFBR (grant No. 18-02-00596). JPUF thanks the DNRF for supporting research at the Cosmic Dawn Center, Niels Bohr Institute, Copenhagen University.

\end{acknowledgements}

\bibliographystyle{aa}
\bibliography{bibliography.bib}

\end{document}